\documentclass[12pt,reqno]{amsart}
\usepackage{hyperref}
\usepackage{appendix}
\usepackage{graphicx}
\usepackage{amscd}
\usepackage{amsmath}
\usepackage{epsfig}
\usepackage{amsfonts}
\usepackage{amssymb}
\usepackage{setspace}
\usepackage{listings}
\usepackage{framed}
\usepackage{xcolor}
\colorlet{shadecolor}{gray!20}

\providecommand{\U}[1]{\protect\rule{.1in}{.1in}}
\providecommand{\U}[1]{\protect\rule{.1in}{.1in}}
\allowdisplaybreaks
\newtheorem{theorem}{Theorem}
\theoremstyle{plain}

\numberwithin{equation}{section}

\input{tcilatex}

\newcommand{\grayitss}[1]{\textcolor{gray}{\textsf{\textit{#1}}}}

\lstdefinestyle{matitalic}{
  language=Mathematica,
  mathescape=true,
  basicstyle=\ttfamily\small,
  columns=fullflexible,
  keepspaces=true,
  breaklines=true,
  upquote=true,
  literate=
    {In[}{{\grayitss{In[}}}1
    {]:=}{{\grayitss{]:=}}}1
    {Out[}{{\grayitss{Out[}}}1
    {]=}{{\grayitss{]=}}}1,
 }

\lstnewenvironment{mat}{\lstset{style=matitalic}}{}

\begin{document}
\title[Quantum Mechanics of Bohr and Sommerfeld]{Old Quantum Mechanics by
Bohr and Sommerfeld \\
from a Modern Perspective}
\author{Kamal Barley}
\address{Department of Mathematics, College of Arts and Sciences, Howard
University, 2141 6th St NW, Washington, DC 200059, U.S.A.}
\email{Kamal.Barley@howard.edu}
\author{Andreas Ruffing}
\address{Landeshauptstadt M\"{u}nchen, Seminar TMG Universit\"{a}t, Referat f\"{u}r Bildung und Sport,
Marienplatz~8, 80331 M\"{u}nchen, Germany}
%LANDESHAUPTSTADT MÜNCHEN, MARIENPLATZ 8, 80331 MÜNCHEN, GERMANY * SEMINAR TMG UNIVERSITÄT, REFERAT FÜR BILDUNG UND SPORT 
\email{alruffing@web.de}
\author{Sergei K. Suslov}
\address{School of Mathematical and Statistical Sciences, Arizona State
University, Tempe, AZ 85287--1804, U.S.A.}
\email{sks@asu.edu}
\date{July 23, 2025} %\today  }
%\dedication{ Dedicated to 100th anniversary of the birth of wave mechanics.}

\begin{abstract}
%We review the classical Bohr model of the atom from the mathematical perspective of wave mechanics.

We review Bohr's atomic model and its extension by Sommerfeld from a mathematical perspective of wave mechanics. The derivation of quantization rules and energy levels is revisited using semiclassical methods. Sommerfeld-type integrals are evaluated by elementary techniques, and connections with the Schr\"{o}dinger and Dirac equations are established. Historical developments and key transitions from classical to quantum theory are discussed to clarify the structure and significance of the old quantum mechanics.
\end{abstract}

\dedicatory{\begin{center}
{\scriptsize{Dedicated to the 100th anniversary of the birth of wave mechanics \cite{SchrQMI}.}}
\end{center}}

\maketitle

\noindent
{\scriptsize{Prediction is very difficult, especially if it's about the future!
%\smallskip
}

\begin{flushright}
\it{Niels Bohr} $\;$
\end{flushright}
}

%\smallskip
\noindent
{\scriptsize{If you want to be a physicist, you must do three things --
first, study mathematics, second, study more mathematics, and third, do the
same. %\smallskip 
}

\begin{flushright}
\it{Arnold Sommerfeld} $\;$
\end{flushright}
}

%%===========================================
%%% Introduction %%%
%%===========================================

\section{Introduction}

The study of blackbody radiation and the quantum theory that emerged from it laid the foundation for Bohr's atomic model, a major step in understanding atomic structure developed about a century ago.
By recognizing the quantum nature of energy and the discrete energy levels of electrons, Planck \cite{Planck}, Einstein \cite{Einstein}, Rutherford \cite{Rutherford}, and Bohr \cite{Bohr1922} helped to explain the behavior of light and matter at the atomic scale, thereby paving the way for the development of quantum mechanics.

Among the primary sources on the so-called \textquotedblleft Old Quantum Mechanics\textquotedblright\ of Bohr and Sommerfeld are the classic publications \cite{AASE, BohrColWorkII, Eckert, Elyashevich1985, Firme, Kragh2012, KraghBohr, MEHRAII, Migdal1985, MilUFN, Reed, Stein, Somm1916, Somm1940, Wilson}, the references therein, and several educational videos~\cite{BohrAtom}. 

{\scshape{Brief history:}} The fine structure of hydrogen atom spectral lines was first observed by Albert~A.~Michelson in 1887 \cite{Michelson, Mich}. After the failure of his ether-wind experiments, he turned to spectroscopy and discovered that the prominent $H_\alpha$ line of the Balmer series was, in fact, a doublet \cite{Biedenharn1983, Kragh1985, Reed}. The electron was discovered by J.~J.~Thomson in 1897 \cite{Thomson}, and Lord Rutherford proposed the planetary model of the atom in 1911. 
Niels Bohr introduced his theory of hydrogen-like systems in 1913 \cite{AASE, BohrColWorkII, Elyashevich1985, KraghBohr, MilUFN, Reed}, and in 1916 Arnold Sommerfeld extended Bohr's quantization rules to the relativistic hydrogen atom \cite{Somm1916} (see also \cite{GranUFN, Kragh1985, SomAS}). An exact solution was finally achieved in 1928 by C.~G.~Darwin \cite{Darwin1928} and W.~Gordon \cite{Gordon1928}, following the discovery of the Dirac equation \cite{Dirac1928, Dirac1977}. Remarkably, their results matched precisely the ‘old’ Sommerfeld formula—an outcome known as the \textquotedblleft Sommerfeld Puzzle\textquotedblright\ \cite{Biedenharn1983}, discussed further in \cite[pp.~426--429]{Eckert}.

In this review article, we aim to explore the following topics from a mathematical perspective:

\noindent 
(i) The Bohr model: circular orbits of electrons in hydrogen-like atoms and the derivation of the Bohr formula (Nobel Prize in Physics, 1922 \cite{Bohr1922}).
\newline
\noindent 
(ii) Wilson and Sommerfeld: quantization rules for multidimensional periodic systems via classical action; derivation of Sommerfeld's relativistic formula for elliptical orbits in classical and wave mechanics.
\newline
\noindent   
(iii) Elementary evaluation of Sommerfeld-type integrals.
\newline
\noindent 
(iv) Additional examples and a resolution of the \textquotedblleft Sommerfeld Puzzle\textquotedblright; a mistake that Schr\"{o}dinger never made.
\newline
\noindent  
(v) Appendix A: Vector calculus tools for uniform circular motion.
\newline
\noindent   
(vi) Appendix B: Instability of the hydrogen atom in classical physics, arising from the electron’s predicted collapse into the nucleus under Rutherford’s model.
\newline
\noindent 
(vii) Appendix C: Independent evaluation of Sommerfeld-type integrals via parameter differentiation.
\newline
\noindent 
(viii) Appendix~D: Letter from Schr\"{o}dinger to Sommerfeld, dated January 29, 1926.
\newline
\noindent 
(ix) Appendix~E: Use of the Mathematica computer algebra system.

Traditional physics textbooks often omit the semiclassical derivation of the Sommerfeld fine structure formula due to its complexity and the fact that an accurate and elegant solution exists within relativistic quantum mechanics. While the semiclassical approach offers conceptual insight, it entails a rigorous and often challenging analysis, rendering it less suitable for introductory courses.

{\scshape{Our goal and motivation:}} 
These notes are intended as a supplement to traditional textbooks \cite{AkhBer, BerLifPit, Dav, La:Lif, Rose, Schiff} and our recent article \cite{Barleyetal2021}, providing original explanations, historical context, and extended discussion on selected topics.
They may serve as a valuable resource in teaching and learning quantum physics, and can support honors projects at any level—from introductory to graduate.
To this end, the presentation is as self-contained as possible.
This work is motivated by a course in the mathematics of quantum mechanics, taught for more than two decades at Arizona State University by one of the authors (SKS) \cite{Barleyetal2021, EEKS-SKS, GorBarSus23, Kryuchkovetal, Kryuch:Sus:Vega12, SusPuzz, Sus:Trey2008, Sus:Vega:Barl} (see also the references therein and \cite{Ruffing}).

%
%%===========================================
%%% Section %%%
%%===========================================
%

\section{Bohr's Atomic Model}

Newton's second law for the uniform circular motion of a charged particle,
such as an electron in the static Coulomb field of a heavy ion with
positive charge $Ze,$ states%
\begin{equation}
ma=F=\frac{Ze^{2}}{r^{2}},\qquad a=\frac{v^{2}}{r}  \label{BM1}
\end{equation}%
by (\ref{A1}) from Appendix~\ref{appendix:a}. (Here, $m\approx 9.\,109\,4\times 10^{-28}\, {\text{%
grams}}$ and $e\approx 4.\,803\,2\times 10^{-10}\, {\text{statcoulombs}}$ are
the electron mass and the absolute value of its electric charge in centimeter-gram-second (cgs) units, respectively.) 
For an electron's linear momentum, $p=mv,$ one obtains%
\begin{equation}
p^{2}=\frac{mZe^{2}}{r}  \label{BM2}
\end{equation}%
and the total energy is given by%
\begin{equation}
E=\frac{p^{2}}{2m}-\frac{Ze^{2}}{r}=-\frac{Ze^{2}}{2r},  \label{BM3}
\end{equation}%
which is exactly one half of the potential energy, as stated by the \textit{virial theorem}.
%

%
%%%%% Figure 1, Bohr's model  %%%%%%
\begin{figure}[hbt!]
\centering
\includegraphics[width=0.5\textwidth]{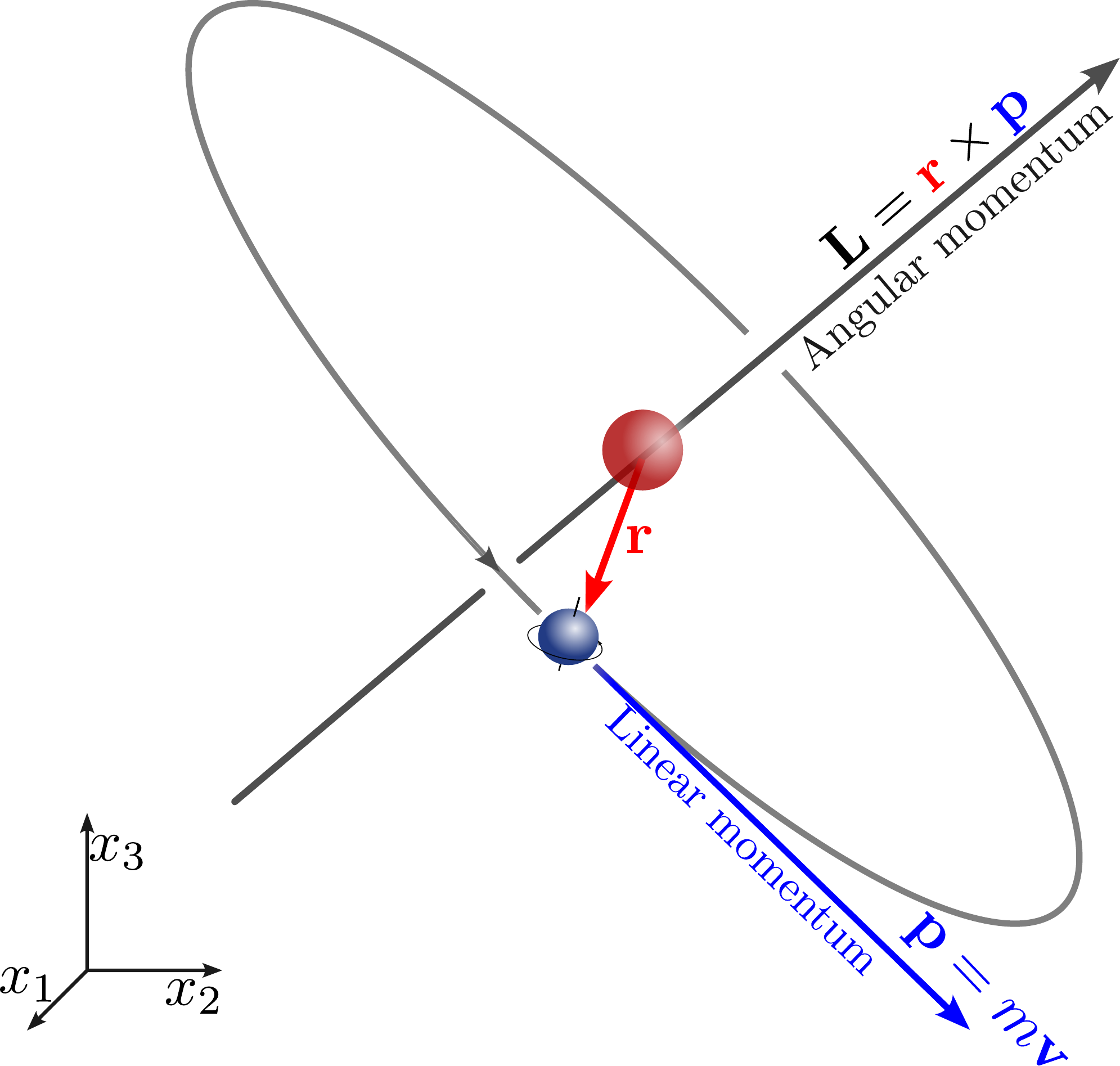}
\caption{Bohr's atom.}
\label{Figure1}
\end{figure}

Niels Bohr \cite{AASE, BohrColWorkII, KraghBohr}, following the experiments of Nicholson \cite{MEHRAI}%
, proposed to quantize the corresponding electron angular momentum%
\begin{equation}
\mathbf{L}=\mathbf{r\times p,\qquad }pr=\hbar n\qquad \left( n=1,2,\ldots
\right)   \label{BM4}
\end{equation}%
in terms of the {\textit{reduced Planck constant}} $\hbar \approx
1.\,054\,6\times 10^{-27}{\, \text{cm}^{2}}{\text{g/s}}$ in cgs units. For 
uniform circular motion, the vectors $\mathbf{r}$ and $\mathbf{p}$ are
perpendicular to each other (\ref{A3}) (see Figure~\ref{Figure1}). As a result, he
derived the so-called Bohr orbits:%
\footnote{%
In terms of the de Broglie wavelength $\lambda ,$ the quantization rule
states that the length of the orbit equals $2\pi r_{n}=n\lambda ,$ where $%
\lambda =h/p=2\pi \hbar /p$.}%
\begin{equation}
r=r_{n}=\frac{\hbar ^{2}n^{2}}{mZe^{2}}  \label{BM5}
\end{equation}
and the corresponding discrete energy levels of the electron:%
\begin{equation}
E_{n}=-\frac{mZ^{2}e^{4}}{2\hbar ^{2}n^{2}},  \label{BM6}
\end{equation}%
where $n=1,2,3,\ldots $ is the \textit{principal quantum number}.

Indeed, using (\ref{BM2}) and (\ref{BM4}): 
\begin{equation}
\frac{mZe^{2}}{r}=p^{2}=\left( \frac{\hbar n}{r}\right) ^{2},  \label{BM7}
\end{equation}%
leads to (\ref{BM5}). Similarly, combining (\ref{BM3}) and (\ref{BM5}):%
\begin{equation}
E_{n}=-\frac{Ze^{2}}{2r_{n}}=-\frac{mZ^{2}e^{4}}{2\hbar ^{2}n^{2}},
\label{BM8}
\end{equation}%
which completes the derivation of Bohr's discrete energy formula (\ref{BM6}).

\begin{figure}[hbt!]
\centering
\includegraphics[width=0.45\textwidth]{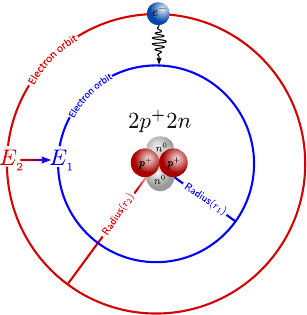}
\caption{Change in electron orbits and energies, $r_2 \to r_1 \, {\rm{and}}\, E_2 \to E_1,$ in a
helium ion $\mathrm{He}^{+}$ upon emission of a photon in Bohr's model. For helium, $Z=2,$ therefore, by (\ref{BM5})--(\ref{BM6}): 
$r_1=.\,264\,6\times 10^{-8}\, \text{cm}, \; r_2=4r_1=1.\,058\,4\times 10^{-8}\,\text{cm}$ and
$E_1= -8.\,719\times 10^{-11}\,\text{erg}=-54.\,424\, \text{eV},\; E_2= E_1/4=-2.\,179\,8\times 10^{-11}\, \text{erg} = -13.\,606\, \text{eV},$ respectively. The emitted photon has a wavelength of $\lambda \approx 30.\,379~{\rm{nm(nanometers)}},$ within the ultraviolet region of the electromagnetic spectrum.}
\label{Figure2}
\end{figure}

In Bohr's atomic model, electrons in the orbits (\ref{BM5}) are stable and do not radiate energy. 
(Instability in the original Rutherford atom is discussed in Appendix~\ref{appendix:b}.) 
Electrons can transition between energy levels (\ref{BM6}) by absorbing or emitting photons (light quanta)
with specific energies. The energy of the photon corresponds to the difference between 
the initial and final energy levels \cite{KraghBohr} (see, for example, Figure~\ref{Figure2}).

%%% Figure 2, changing orbits in helium atom  %%%%
%%%%%%%%%%%%%%%%%%%%%%%%%%%%%%%%%%%%%%%%%%%

Mendeleev's Periodic Table and the Bohr model are two significant
developments in the understanding of the structure of atoms and elements.
Mendeleev's table, introduced in 1869, organized elements by increasing atomic
weight and recurring chemical properties \cite[pp.~2--3]{Stein}. The Bohr model, proposed in 1913 \cite{BohrColWorkII},
offered a theoretical framework for atomic structure by depicting electrons orbiting
the nucleus in fixed, quantized energy levels.
This simplified representation of electronic structure directly underpins
the periodic arrangement of elements in the modern table \cite{SomAS}.

%
%%===========================================
%%% Section %%%
%%===========================================
%

\section{Wilson and Sommerfeld Quantization Rules in Wave Mechanics}

{\scshape{Topics to review:}} 
Kepler problems in classical mechanics \cite{Biedenharn1983}, \cite[pp.~146--148]{Gantmacher}, \cite[pp.~92--102, pp.~466--477, pp.~481--482]{Goldstein}, \cite[pp.~84--90, pp.~109--119, pp.~251--258]{SomAS};
spherical harmonics \cite{EEKS-SKS, La:Lif, Ni:Uv, Varshalovich1988};
the Schr\"{o}dinger equation \cite{Dav, La:Lif, MerzB, Schiff}, relativistic Schr\"{o}dinger 
and Dirac equations \cite{AkhBer, Barleyetal2021, BerLifPit, Dav, Schiff, Sus:Vega:Barl};
the spinor spherical harmonics \cite{AkhBer, BerLifPit, Rose, Sus:Trey2008, Sus:Vega:Barl, Varshalovich1988}, 
and separation of variables for the Dirac equation in a central field; 
semiclassical approximation \cite{BerryMount-SKS, Ghatetal, Ni:Uv, Schiff}.

{\scshape{Brief history:}} As an extension of Bohr's rules, Wilson \cite{Wilson} and Sommerfeld \cite{Somm1916} 
independently proposed a method for quantizing action integrals in classical mechanics 
for a multidimensional periodic system over one period of motion (see also \cite[footnote on p.~111]{SomAS}).

\subsection{Sommerfeld Fine Structure Formula}

We follow \cite{Biedenharn1983} and \cite[pp.~251--258]{SomAS}, with some differences in presentation.
The classical relativistic Hamiltonian, or total energy $E$, of hydrogen-like systems under an
attractive Coulomb potential, has the quadratic form:%
\begin{equation}
\left( E+\frac{Ze^{2}}{r}\right) ^{2}=\mathbf{p}^{2}c^{2}+m^{2}c^{4}.
\label{BidSom1}
\end{equation}%
In polar coordinates,%
\begin{equation}
\mathbf{p}^{2}=\left( p_{r}\right) ^{2}+\frac{1}{r^2}\left( p_{\theta }\right)
^{2},  \label{BidSom2}
\end{equation}%
with $p_{r}=\gamma m\overset{\cdot}{r}=\gamma m\ (dr/dt)$ (radial momentum), and 
$p_{\theta }=\gamma mr^{2}\overset{\cdot}{\theta }=\gamma mr^{2}\ (d\theta /dt)$
(angular momentum). [In the relativistic case, $\gamma =\left(
1-v^{2}/c^{2}\right) ^{-1/2}$ is the familiar Lorentz factor.]

Due to conservation of angular momentum, $p_{\theta }$ is constant. Introducing
the new variable $s=1/r$, we note that%
\begin{equation}
\frac{ds}{d\theta }=-\frac{p_{r}}{p_{\theta }}.  \label{BidSom3}
\end{equation}%
In this notation, Eq.~(\ref{BidSom1}) becomes%
\begin{equation}
\left( \frac{E}{mc^{2}}+\frac{Ze^{2}}{mc^{2}}\, s\right)^{2}=1+\left( \frac{%
p_{\theta }}{mc}\right) ^{2}\left[ \left( \frac{ds}{d\theta }\right)
^{2}+s^{2}\right] .  \label{BidSom4}
\end{equation}%
Differentiation with respect to $\theta$ yields the linear
ordinary differential equation:%
\begin{equation}
\frac{d^{2}s}{d\theta ^{2}}+\omega ^{2}\left( s-D\right) =0.  \label{BidSom5}
\end{equation}%
Here, by definition,%
\begin{equation}
\omega ^{2}=1-\frac{Z^{2}e^{4}}{c^{2}p_{\theta }^{2}},\qquad D=\frac{Ze^{2}E%
}{\omega ^{2}c^{2}p_{\theta }^{2}}.  \label{BidSom6}
\end{equation}

Solving (\ref{BidSom5}) yields the relativistic Kepler orbits in the form%
\begin{equation}
s=\frac{1}{r}=C_{1}\cos \left( \omega \theta \right) +C_{2}\sin \left(
\omega \theta \right) +D,  \label{BidSom7}
\end{equation}%
where $C_{1}$ and $C_{2}$ are constants. If the point of closest approach
(perihelion) occurs at $\theta =0$, then $C_{2}=0$, and%
\begin{equation}
\frac{1}{r}=C_{1}\cos \left( \omega \theta \right) +D.  \label{BidSom8}
\end{equation}

\begin{figure}[hbt!]
\centering
\includegraphics[width=0.5\textwidth]{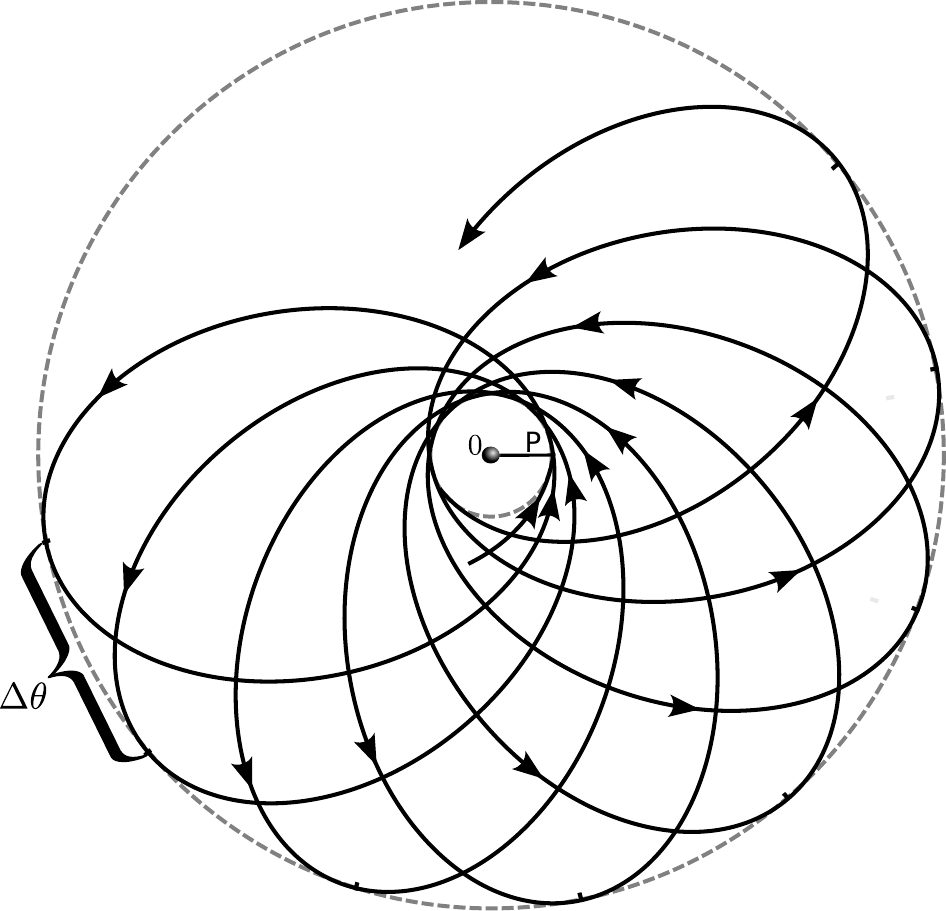}
\caption{Relativistic Kepler motion \cite[p.~254]{SomAS}. 
(Here, $\textrm{O}$ is the fixed focus at which the nucleus is situated;
$\textrm{P}$ is the initial position of the perihelion.)
The perihelion and aphelion move along two concentric
circles around the nucleus at $\textrm{O}$.}
\label{FigureRelativistic}
\end{figure}

Classical relativistic Kepler orbits have the form of conic sections, as in
the nonrelativistic case \cite{Biedenharn1983, SomAS}, but with a new angular variable $\phi =\omega \theta.$ Thus, for
elliptical orbits (bound states), the motion from one perihelion ($\phi =0$) to
the next ($\phi =2\pi$) requires $\theta =2\pi /\omega \,$, with a per-revolution shift of $\Delta\theta =2\pi /\omega - 2\pi$
(see Figure~\ref{FigureRelativistic}).
%

%%%%% Figure 3: Ellipses, Kepler's problem  %%%%%%

Introducing the eccentricity $\epsilon$, we have, for $\phi = 0$, the perihelion
distance $r_{\textrm{min}} = a(1 - \epsilon)$, and for $\phi = \pi$, the aphelion distance 
$r_{\textrm{max}} = a(1 + \epsilon)$. In standard geometrical terms, the orbit equation is%
\begin{equation}
\frac{1}{r} = \frac{1 + \epsilon \cos \left( \omega \theta \right)}{a\left(
1 - \epsilon ^{2} \right)}.\label{BidSom9}
\end{equation}%
One can now apply the original Wilson--Sommerfeld quantization rules:%
\begin{equation}
\int_{\theta = 0}^{\theta = 2\pi} p_{\theta}\, d\theta = h\, n_{\theta} \qquad
\left( \text{which gives } p_{\theta} = \hbar\, n_{\theta} \right) \label%
{BidSom10}
\end{equation}%
and%
\begin{equation}
\int_{\theta = 0}^{\theta = 2\pi/\omega} p_{r}\, dr = h\, n_{r}. \label{BidSom11}
\end{equation}

To evaluate the last integral, we transform the radial momentum as follows:%
\begin{equation}
p_{r} = \gamma m \overset{\cdot}{r} = \gamma m \left( \frac{dr}{d\theta} \right) 
\overset{\cdot}{\theta} = \frac{p_{\theta}}{r^{2}} \left( \frac{dr}{d\theta} \right). \label{BidSom12}
\end{equation}%
Thus,%
\begin{equation}
p_{r}\, dr = p_{\theta} \left( \frac{1}{r} \frac{dr}{d\theta} \right)^{2} d\theta 
= p_{\theta} \epsilon^{2} \omega \frac{\sin^{2} \phi}{\left( 1 + \epsilon \cos \phi \right)^{2}}\, d\phi \label{BidSom13}
\end{equation}%
upon using the orbit equation (\ref{BidSom9}). The radial quantization condition (\ref{BidSom11}) then becomes%
\begin{equation}
\frac{1}{2\pi} \int_{\phi = 0}^{\phi = 2\pi} \frac{\epsilon^{2} \sin^{2} \phi\,
d\phi}{\left( 1 + \epsilon \cos \phi \right)^{2}} = \frac{n_{r}}{\omega n_{\theta}}. \label{BidSom14}
\end{equation}%
For evaluation of the integral,%
\begin{equation}
\frac{1}{2\pi} \int_{\phi = 0}^{\phi = 2\pi} \frac{\epsilon^{2} \sin^{2} \phi\,
d\phi}{\left( 1 + \epsilon \cos \phi \right)^{2}} = \left( 1 - \epsilon^{2} \right)^{-1/2} - 1, \label{BidSom15}
\end{equation}%
see \cite[pp.~476--477, German edn.]{SomAS} and our complementary Mathematica notebook \cite{BarleySusMath}.

As a result, we obtain%
\begin{equation}
\frac{1}{1 - \epsilon^2} = \left( 1 + \frac{n_{r}}{\omega n_{\theta}} \right)^2 \label{BidSom16}
\end{equation}%
and%
\begin{equation}
D = \frac{\alpha Z E}{n_{\theta}^{2} \omega^{2} \hbar c} = \frac{1}{a \left( 1 - \epsilon^2 \right)} 
\quad \left( \alpha = \frac{e^{2}}{\hbar c} \; \textrm{is the fine-structure constant} \right). \label{BidSom17}
\end{equation}%
Finally, using the last two equations together with the orbit (\ref{BidSom9}) and energy (\ref{BidSom4})
equations, after tedious but straightforward calculations, one arrives at
the original Sommerfeld formula:%
\begin{equation}
\frac{E_{n_{r}, n_{\theta}}}{mc^{2}} = \left( 1 + \frac{\alpha^2 Z^2}{\left(
n_{r} + \left( n_{\theta}^{2} - \alpha^2 Z^2 \right)^{1/2} \right)^2} \right)^{-1/2}, \label{BidSomEnd}
\end{equation}%
where $n_{r}$ (the {\textit{radial quantum number}}) and $n_{\theta}$ (the {\textit{azimuthal quantum number}}) are positive integers. This result made it possible to explain, for the first time, the fine structure of spectral lines.
(For further details, see \cite{Biedenharn1983}, \cite[pp.~251--258]{SomAS}, and Appendix~\ref{appendix:e}.)
\noindent {\textbf{Note.}} Equations (\ref{BidSom6}), (\ref{BidSom10}), and (\ref{BidSom16})--(\ref{BidSomEnd}) allow us to determine the quantized values of parameters of the electron’s elliptical orbits (\ref{BidSom9}) as follows:%
\begin{eqnarray}
\omega_{n_{\theta}} n_{\theta} &=& \left( n_{\theta}^{2} - \alpha^{2} Z^{2} \right)^{1/2}, \label{BidSom18} \\
\epsilon_{n_{r}, n_{\theta}} &=& \sqrt{n_{r}} \cdot 
\frac{\left( n_{r} + 2 \sqrt{n_{\theta}^{2} - \alpha^{2} Z^{2}} \right)^{1/2}}{n_{r} + \sqrt{n_{\theta}^{2} - \alpha^{2} Z^{2}}}, \label{BidSom19} \\
a_{n_{r}, n_{\theta}} &=& \frac{a_{0}}{Z} \left( n_{r} + \sqrt{n_{\theta}^{2} - \alpha^{2} Z^{2}} \right) \label{BidSom20} \\
&&\times \sqrt{\alpha^{2} Z^{2} + \left( n_{r} + \sqrt{n_{\theta}^{2} - \alpha^{2} Z^{2}} \right)^{2}}, \notag
\end{eqnarray}%
where $a_{0} = \hbar^{2} / (m e^{2})$ is the familiar Bohr radius. These formulas generalize the circular orbits. (Further details can be found in the complementary Mathematica notebook \cite{BarleySusMath} and in Appendix~\ref{appendix:e}.)%
\footnote{Classical solutions of the relativistic Kepler problem are also discussed in \cite[pp.~481--482]{Goldstein} and \cite[pp.~100--102]{LaLif2}.}

\begin{figure}[hbt!]
\centering
\includegraphics[width=0.5\textwidth]{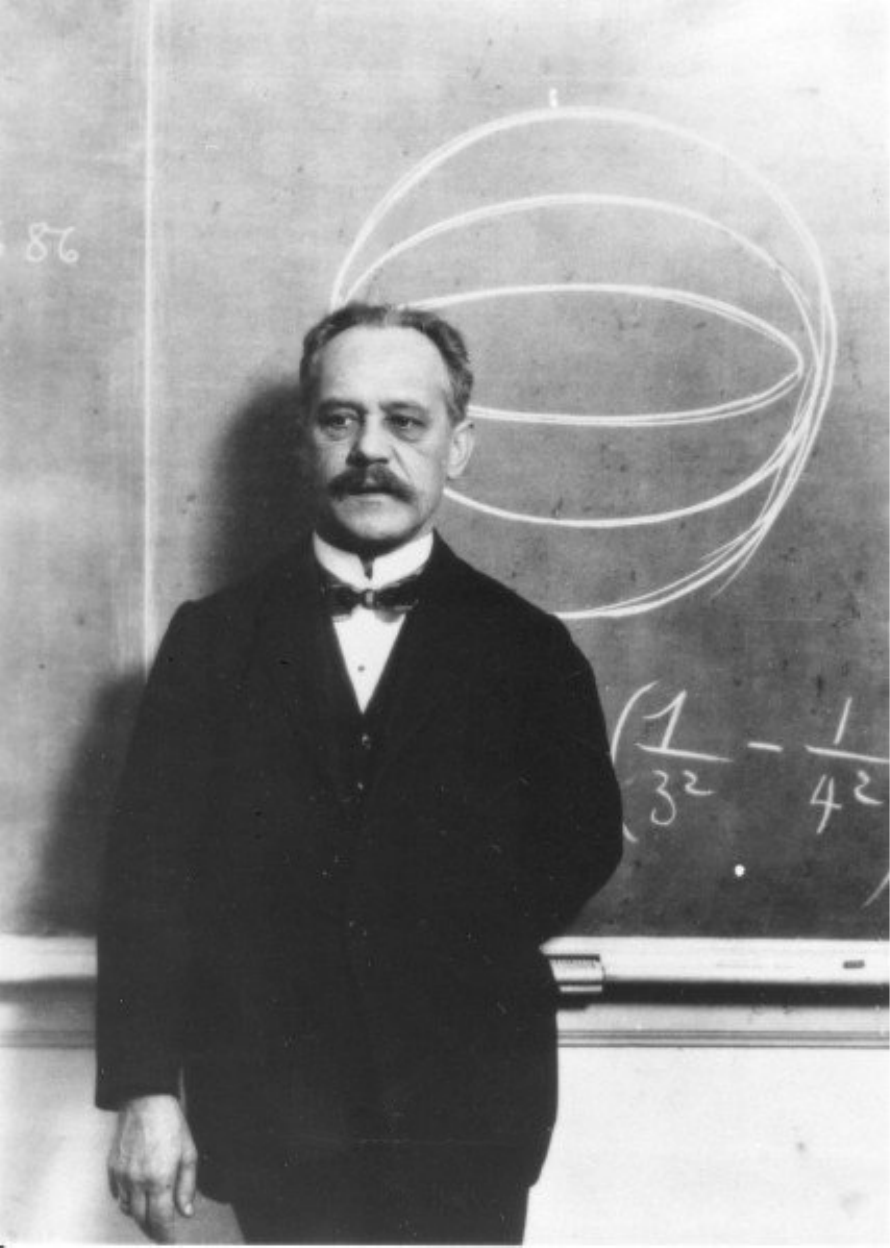}
\caption{Sommerfeld's elliptic orbits for the hydrogen atom \cite[p.~116]{SomAS}.}
\label{SommerfeldEllipses}
\end{figure}

In the nonrelativistic limit, we obtain%
\begin{eqnarray}
\omega_{n_{\theta}} &=& 1 - \frac{\alpha^{2} Z^{2}}{2 n_{\theta}^{2}} - \frac{\alpha^{4} Z^{4}}{8 n_{\theta}^{4}} + \mathcal{O} \left( \alpha^{6} \right), \label{BidSom21} \\
\epsilon_{n_{r}, n_{\theta}} &=& \frac{\sqrt{n_{r}} \left( n_{r} + 2 n_{\theta} \right)^{1/2}}{n_{r} + n_{\theta}} 
+ \frac{\sqrt{n_{r}} \alpha^{2} Z^{2}}{2 \left( n_{r} + n_{\theta} \right)^{2} \left( n_{r} + 2 n_{\theta} \right)^{1/2}} \label{BidSom22} \\
&& + \frac{\sqrt{n_{r}} \left( 3 n_{r} + 5 n_{\theta} \right) \alpha^{4} Z^{4}}{8 n_{\theta} \left( n_{r} + n_{\theta} \right)^{3} \left( n_{r} + 2 n_{\theta} \right)^{3/2}} + \mathcal{O} \left( \alpha^{6} \right), \notag \\
\dfrac{Z a_{n_{r}, n_{\theta}}}{a_{0}} &=& \left( n_{r} + n_{\theta} \right)^{2} 
- \frac{\alpha^{2} Z^{2} \left( 2 n_{r} + n_{\theta} \right)}{2 n_{\theta}} \label{BidSom23} \\
&& - \frac{\alpha^{4} Z^{4}}{8} \left( \frac{2 n_{r}}{n_{\theta}^{3}} + \frac{1}{\left( n_{r} + n_{\theta} \right)^{2}} \right) + \mathcal{O} \left( \alpha^{6} \right), \notag
\end{eqnarray}%
and%
\begin{equation}
\frac{E_{n_{r}, n_{\theta}}}{mc^{2}} = 1 - \frac{\alpha^{2} Z^{2}}{2 \left( n_{r} + n_{\theta} \right)^{2}} 
- \frac{\alpha^{4} Z^{4} \left( 4 n_{r} + n_{\theta} \right)}{8 n_{\theta} \left( n_{r} + n_{\theta} \right)^{4}} 
+ \mathcal{O} \left( \alpha^{6} \right), \label{BidSom24}
\end{equation}%
where $\alpha = e^{2} / (\hbar c)$ and $c \rightarrow \infty$ (see our Mathematica file). [This asymptotic expansion will be analyzed below (\ref{lim3}).]
In this limit, we recover Sommerfeld's elliptic orbits for hydrogen-like systems \cite[pp.~109--119]{SomAS} (see Figure~\ref{SommerfeldEllipses}).
As follows from (\ref{BidSom22}), Bohr's circular orbits occur only when $n_{r} = 0$. 

%%%%% Figure 4: Sommerfeld's Ellipses, Kepler's problem  %%%%%%

In his classical book, a ``Bible" of Atomic Theory, Sommerfeld concludes \cite[p.~258]{SomAS}:%
{\textit{
The above-calculated {$\textbf{energy levels}$} and the line complexes to be
derived from them also retain their validity in wave mechanics. The
way in which wave mechanics derives them is not only far less picturesque
{$($\textbf{anschaulich}$)$}, but also much more laborious than the way described
above. Hence it was necessary to carry out the calculation as far as
possible according to the method of the older quantum theory; the
inferences drawn can then later be taken over directly into wave mechanics.}}

Indeed, in his introductory book on wave mechanics, Sommerfeld also discussed the quantization of Kepler's problems for the relativistic Schr\"{o}dinger 
and Dirac equations \cite[pp.~100--104, 112--118, and 282--286]{SomQM}.
(See also \cite{Somm1940}.)

\subsection{Fine structure formula in wave mechanics}

In the development of quantum theory, the Bohr--(Wilson)--Sommerfeld
quantization rule served as an original “bridge” between classical and quantum
mechanics (for historical details, see \cite{MEHRAI}, \cite{MEHRAII}, \cite{MilUFN}, \cite{Reed}, \cite%
{Somm1916}, \cite{SomAS}, and \cite{Wilson}).
Today, the Schr\"{o}dinger \cite{Schroedinger} and Dirac \cite{Dirac} wave equations 
are used to analyze the corresponding Kepler problems.
How did Schr\"{o}dinger derive his celebrated equation and subsequently apply it to the hydrogen atom?
According to his own testimony \cite{SchrQMI, SchrodingerCohrent, SchrPhysRev1926} and \cite[030\dag\;~pp.~141--143]{Meyenn2011},%
{\footnote{In his letter to Einstein dated November~3, 1925, he writes: %
\textit{A few days ago, I read with great interest the ingenious theses of Louis de Broglie, which I finally got hold of ...\;.}
}}
de Broglie’s seminal work on the wave theory of matter (1923–24) \cite{deBroglie} and Einstein’s studies on ideal Bose gases (1924–25) 
laid the foundation for the discovery of wave mechanics (see also \cite{Barleyetal2021}, \cite{MEHRAIII}, and \cite{Moore1989}).
The phenomenological quantization rules of the “old” quantum theory \cite%
{Somm1916, Wilson} are, in modern physics, derived from
the corresponding wave equations via the so-called semiclassical approximation—
also known as the Wentzel–Kramers–Brillouin (WKB) method—
as developed in \cite{BerryMount-SKS}, \cite{Brillouin}, \cite{Kramers}, \cite{La:Lif}, \cite%
{Ni:Uv}, and \cite{Wentzel}.

This approximation refines the Bohr--Sommerfeld quantization rule within the framework of wave mechanics.
The WKB method, which yields approximate solutions to wave equations, 
leads to a quantization condition similar to the Bohr--Sommerfeld rule, 
but with a crucial phase correction.

During separation of variables in spherical coordinates, the quantization of angular momentum and spin is exact, 
since the concept of spin is inherently built into the structure 
of the wave equation itself \cite{Kryuchkovetal}.
After this, the analysis reduces to solving radial equations.
We follow \cite[pp.~95--96]{Barleyetal2021} with somewhat different details.
Let us recall the one-dimensional stationary Schr\"{o}dinger equation:%
\begin{equation}
u^{\prime \prime }+\frac{2m}{\hbar ^{2}}\left[ E-U\left( x\right) \right] u=0.%
\label{SchEq}
\end{equation}
For a particle in a central field, the corresponding $3D$ wave equations can be separated in spherical
coordinates, yielding a radial equation of the form:
\begin{equation}
u^{\prime \prime }(x)+q(x)\,u(x)=0,  \label{wkb1}
\end{equation}
where $x^{2}q(x)$ is continuous along with its first and second
derivatives for $0 \leq x \leq b < \infty$.  
These equations can be approximately solved by the WKB method.

{\scshape{More to review:}} 
The WKB wave functions, their relation to Airy functions \cite{NIST}, associated quantization rules, 
and further technical details are discussed in \cite{Ni:Uv, Schiff}, and elsewhere.
We recommend reviewing Sections $\S$19, pp.~235--251, and $\S$28, pp.~178--188, 
on the semiclassical approximation in \cite{Ni:Uv, Schiff}, as well as \cite[pp.~380--390]{NikNovUv}, 
and Chapter~9 of \cite{NIST} on Airy functions.

It is well known that the traditional semiclassical approximation breaks down
near $x=0$ for central fields. However, using the change of variables $x = e^z$ and $u = e^{z/2} v(z)$ 
transforms the equation into the new form:
\begin{equation}
v^{\prime \prime }(z)+q_{1}(z)\,v(z)=0,  \label{wkb2}
\end{equation}
where 
\begin{equation}
q_{1}(z)=-\dfrac{1}{4}+\left( x^{2}q(x)\right)_{x=e^{z}}.  \label{wkb3}
\end{equation}
This is known as Langer's modification \cite{BerryMount-SKS, Langer1931, Langer1937}. As $% 
z\rightarrow -\infty$ (i.e., $x\rightarrow 0$), the function $q_{1}(z)$
approaches the constant:
\begin{equation*}
-1/4+\lim_{x\rightarrow 0}x^{2}q(x), \qquad \text{and} \quad
\lim_{z\rightarrow -\infty }q_{1}^{(k)}(z)=0\quad (k=1,2).
\end{equation*}
Thus, $q_{1}(z)$ and its derivatives vary slowly for large negative $z$ \cite[p.~387]{NikNovUv}.

The WKB method can then be applied to this transformed equation, 
and in the original equation one replaces $q(x)$ with:
\begin{equation}
q(x)-\dfrac{1}{4x^{2}} = p^{2}_{\text{effective}}(x)  \label{wkb3a}
\end{equation}
(see \cite{BerryMount-SKS}, \cite{Langer1931}, \cite{Langer1937}, and \cite{Ni:Uv} for further details).

The Bohr--Sommerfeld quantization rule, derived for example in \cite{Ni:Uv} 
and \cite{Schiff}, takes the form:
\begin{equation}
\int_{r_{1}}^{r_{2}}p(r)\,dr = \pi \left( n_{r} + \dfrac{1}{2} \right) \qquad
(n_{r}=0,1,2,\dots \; \text{radial quantum number})  \label{wkbBoSom}
\end{equation}
provided $p(r_{1})=p(r_{2})=0$.

For all Coulomb problems under consideration, we utilize a generic integral originally
evaluated by Sommerfeld \cite[pp.~611--612]{SomAS} using complex analysis: if 
\begin{equation}
p(r) = \sqrt{ -A + \dfrac{B}{r} - \dfrac{C}{r^{2}} } \qquad  \left(A,C > 0\right) ,  \label{SomABC}
\end{equation}
then:
\begin{equation}
\int_{r_{1}}^{r_{2}}p(r)\,dr = \pi \left( \dfrac{B}{2\sqrt{A}} - \sqrt{C} \right)  \label{SomABCInt}
\end{equation}
with $p(r_{1}) = p(r_{2}) = 0$
(see also \cite[pp.~468--470]{Goldstein}). In Section~\ref{sec:SommerfeldIntegral} and Appendix~\ref{appendix:c}, 
we present two independent elementary evaluations of this integral.

As a result, for the discrete energy levels, we obtain the following generic equation:
\begin{equation}
\dfrac{B}{2\sqrt{A}} - \sqrt{C} = n_{r} + \dfrac{1}{2}  , \label{wkbBoSomSus}
\end{equation}
which is valid for all Coulomb-type problems under consideration and beyond \cite{SomQM, SusPuzz}.

\noindent {\textbf{Kepler problems in wave mechanics.}}  
For the well-known case of the {\it{non-relativistic Coulomb problem,}} the radial equation in dimensionless units reads \cite{La:Lif, Schiff, SchrQMI}:
\begin{eqnarray}
&&u^{\prime \prime }+\left[ 2\left( \varepsilon _{0}+\frac{Z}{x}\right) -%
\frac{l\left( l+1\right)}{x^{2}}\right] u=0  \label{wkb5.4a} \\
&&\left( \varepsilon _{0}=\dfrac{E}{E_{0}},\,\, E_{0}=\dfrac{e^{2}}{%
a_{0}},\,\, a_{0}=\dfrac{\hbar ^{2}}{me^{2}},\,\, x=\dfrac{r}{a_0}\right),  \notag
\end{eqnarray}%
where $l=0,1,2,\dots$ is the quantized orbital angular momentum.

In applying the Bohr--Sommerfeld quantization rule, one must, in accordance with (\ref{wkb3a}), use:
\begin{equation}
p(r) = \left[ 2\left( \varepsilon _{0} + \frac{Z}{r} \right) - \frac{(l + 1/2)^2}{r^2} \right]^{1/2}, \qquad p(r_1) = p(r_2) = 0, \label{wkb5.6}
\end{equation}
as corrected by Langer’s substitution, discussed, for instance, in \cite{Barleyetal2021},  \cite{BerryMount-SKS}, \cite{Ni:Uv}, and \cite{SusPuzz}.

Identifying parameters in the generic integral \eqref{SomABC}, we have:
\[
A = -2\varepsilon_0, \qquad B = 2Z, \qquad C = (l + 1/2)^2.
\]
Substituting into the quantization rule \eqref{wkbBoSomSus}, we find:
\begin{equation}
\dfrac{Z}{\sqrt{-2\varepsilon_0}} - l - \dfrac{1}{2} = n_r + \dfrac{1}{2}. \label{wkb5.8}
\end{equation}
Solving for $\varepsilon_0$ yields the exact energy levels for the non-relativistic hydrogen-like atom:
\begin{equation}
\varepsilon_0 = \dfrac{E}{E_0} = -\dfrac{Z^2}{2(n_r + l + 1)^2}. \label{wkb5.9}
\end{equation}
Here, $n = n_r + l + 1$ is the \textit{principal quantum number}, recovering Bohr’s formula for discrete energy levels, as presented in Eq.~\eqref{BM6} via the semiclassical approximation.
(The WKB method is typically introduced in quantum mechanics only after the exact solution has been established \cite{La:Lif}.)
%

%%%%%% Figure 5 %%%%%%
%
\begin{figure}[hbt!]
\centering
\includegraphics[width=0.715\textwidth]{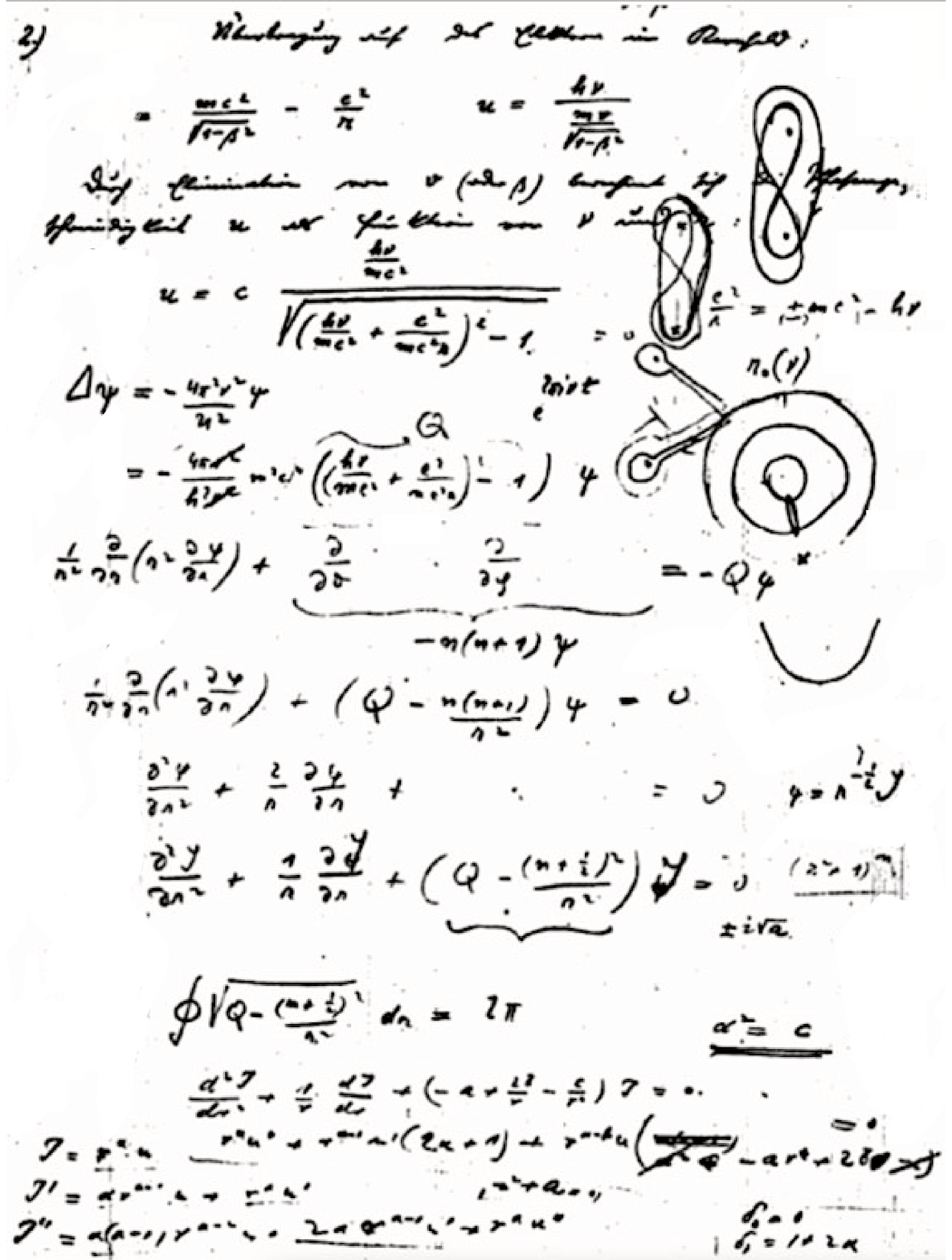}
\caption{A page from Notebook $N1,$ showing the first record of the wave equation \cite{Kragh1982, Kragh1984, Moore1989, Stein} (likely written around Christmas 1925).}
%( Schr\"{o}dinger's notebooks are reproduced in the Archive for the History of Quantum Physics (AHQP); for more details, see  \cite{Joas2009}, \cite{MEHRAIII}.)}
\label{Figure3}
\end{figure}

Our main goal is to analyze the corresponding relativistic problems. For the {\it relativistic Schr\"{o}dinger equation}, one writes \cite{Barleyetal2021}:
\begin{equation}
u^{\prime \prime }+\left[ \left( \varepsilon +\frac{\mu }{x}\right) ^{2}-1-%
\frac{l(l+1)}{x^{2}}\right] u = 0,  \label{wkb5.10}
\end{equation}
(see Figure~\ref{Figure3} for the original version%
\footnote{Schr\"{o}dinger's notebooks are reproduced in the Archive for the History of Quantum Physics (AHQP); for further details, see  \cite{Joas2009, MEHRAIII}.}%
), and applies Langer's transformation to define the effective momentum:
\begin{equation}
p(x) = \left[ \left( \varepsilon +\frac{\mu }{x} \right)^2 - 1 - \frac{(l+1/2)^2}{x^2} \right]^{1/2}. \label{wkb5.11}
\end{equation}
We identify the parameters:
\[
A = 1 - \varepsilon^2, \qquad B = 2\mu \varepsilon, \qquad C = (l + 1/2)^2 - \mu^2.
\]
Applying the Bohr--Sommerfeld quantization condition \eqref{wkbBoSomSus} yields:
\begin{equation}
\dfrac{\mu \varepsilon}{\sqrt{1 - \varepsilon^2}} = n_r + \nu + 1, \qquad \varepsilon = \dfrac{E}{mc^2}, \label{wkb5.12}
\end{equation}
which gives the exact relativistic energy levels:
\begin{equation}
E = E_{n_r} = \frac{mc^2}{\sqrt{1 + \left( \dfrac{\mu}{n_r + \nu + 1} \right)^2}},  \qquad \left(n_r = 0, 1, 2, \dots\right), \label{sol16}
\end{equation}
where 
\begin{equation}
\mu = \frac{Ze^2}{\hbar c}, \qquad \nu = \nu_{\text{Schr\"{o}dinger}} = -\frac{1}{2} + \sqrt{ \left( l + \frac{1}{2} \right)^2 - \mu^2 }. \label{sol17}
\end{equation}

In the non-relativistic limit $c \rightarrow \infty$ (or $\mu \rightarrow 0$), one obtains \cite{Dav, Schiff}:
\begin{eqnarray}
\frac{E_{n_r,\,l}}{mc^2} &=& \frac{1}{\sqrt{ 1 + \dfrac{\mu^2}{\left[ n_r + \frac{1}{2} + \sqrt{(l + \frac{1}{2})^2 - \mu^2} \right]^2 }}}  \notag \\
&=& 1 - \frac{\mu^2}{2n^2} - \frac{\mu^4}{2n^4} \left( \frac{n}{l + 1/2} - \frac{3}{4} \right) + \mathcal{O}(\mu^6), \quad \mu \rightarrow 0, \label{lim2}
\end{eqnarray}
where $n = n_r + l + 1$ is the corresponding non-relativistic principal quantum number (see also \cite{BarleySusMath} for a complementary Mathematica notebook).

In this Taylor expansion:
\begin{itemize}
\item The first term corresponds to the rest energy $E_0 = mc^2$.
\item The second term yields the non-relativistic Schr\"{o}dinger energy eigenvalue.
\item The third term represents the \textit{fine structure}, lifting the degeneracy among states with the same $n$ but different $l$.
\end{itemize}
Sommerfeld's fine structure formula for the relativistic Coulomb problem represents one of the most significant achievements 
of the `old' quantum mechanics \cite[pp.~251--258]{SomAS}.
Here, we derive this result in the semiclassical approximation using the
{\it{radial Dirac equations}} (separation of variables in spherical coordinates is discussed
in detail in Refs.~\cite{Bethe:Sal}, \cite{Ni:Uv}, \cite{Rose}, \cite{Sus:Trey2008}, and \cite{Sus:Vega:Barl}).

In dimensionless units, one of the second-order differential equations for the Dirac spinor component takes the form:
\begin{equation}
v_{1}^{\prime \prime} + \frac{(\varepsilon^{2} - 1)x^{2} + 2\varepsilon \mu x - \nu(\nu + 1)}{x^{2}}\,v_{1} = 0, \label{wkb5.13}
\end{equation}
while the second equation is obtained via the substitution $\nu \rightarrow -\nu$ (see Eqs.~(6.58)--(6.59) in Ref.~\cite{Sus:Trey2008} and/or Eqs.~(3.81)--(3.82) in Ref.~\cite{Sus:Vega:Barl}).

\noindent\textbf{Note.}
It should be emphasized that the above (Schr\"{o}dinger-type) form of the radial equations—veri\-fiable using a computer algebra system \cite{EEKS-SKS}—is critical for the successful application of the WKB approximation to the relativistic Coulomb problem \cite{Barleyetal2021}.
(For an alternative, see, for example, \cite{Good1953, Lu1970}.)

Applying Langer's transformation leads to the effective momentum function:
\begin{equation}
p(x) = \left[ \left( \varepsilon + \frac{\mu}{x} \right)^2 - 1 - \frac{(\nu + 1/2)^2 + \mu^2}{x^2} \right]^{1/2}. \label{wkb5.14}
\end{equation}
Thus, for the Dirac equation, we identify:
\[
A = 1 - \varepsilon^2, \qquad B = 2\mu \varepsilon, \qquad C = (\nu + 1/2)^2.
\]
Applying the Bohr--Sommerfeld quantization rule \eqref{wkbBoSomSus}, we again arrive at Eq.~\eqref{wkb5.12},
yielding the corresponding energy spectrum:
\begin{equation}
E = E_{n_r,\, j} = \frac{mc^2}{\sqrt{1 + \dfrac{\mu^2}{(n_r + \nu)^2}}}, \qquad (n_r = 0, 1, 2, \dots), \label{SommDir}
\end{equation}
with the adjustment $n_r \rightarrow n_r - 1$ as discussed in \cite{Ni:Uv, Sus:Trey2008, Sus:Vega:Barl}.
Here, once again, $\mu = Ze^2 / (\hbar c)$, and in Dirac theory,
\begin{equation}
\nu = \nu_{\text{Dirac}} = \sqrt{(j + 1/2)^2 - \mu^2}, \label{SommDirNu}
\end{equation}
where $j = 1/2, 3/2, 5/2, \dots$ is the total angular momentum (including spin). 
[Observe that only at this stage, in the `old' formula (\ref{BidSomEnd}), can we identify 
Sommerfeld’s {\it azimuthal quantum number}, $n_{\theta}$, as $n_{\theta} = j + 1/2$.]

In the non-relativistic limit ($\mu \rightarrow 0$), the Dirac--Sommerfeld formula yields \cite{Bethe:Sal, Dav, Schiff, Sus:Vega:Barl}:
\begin{equation}
\frac{E_{n_r,\, j}}{mc^2} = 1 - \frac{\mu^2}{2n^2} - \frac{\mu^4}{2n^4} \left( \frac{n}{j + 1/2} - \frac{3}{4} \right) + \mathcal{O}(\mu^6), \label{lim3}
\end{equation}
where $n = n_r + j + 1/2$ is the principal quantum number for hydrogen-like atoms (see also \cite{BarleySusMath}).

In this expansion:
\begin{itemize}
\item The first term is the rest mass energy of the electron, $E_0 = mc^2$.
\item The second term recovers the non-relativistic Schrödinger energy.
\item The third term provides the fine-structure correction, originating from spin–orbit interaction in the Pauli approximation.
\end{itemize}

This prediction agrees with experimental data on fine-structure splitting in hydrogen-like systems.
In contrast, Schr\"{o}dinger’s relativistic formulation fails to accurately describe the fine structure of hydrogen-like atoms (e.g., hydrogen, ionized helium, doubly ionized lithium). For example, the total fine-structure splitting at $n = 2$ is overestimated by a factor of $8/3$ relative to Sommerfeld’s prediction, which is consistent with experimental measurements.

Indeed, the maximum spread in the fine-structure levels occurs for $l = 0$ and $l = n - 1$, with total angular momentum $j = 1/2$ and $j = n - 1/2$, respectively, as seen in Eqs.~\eqref{lim2} and~\eqref{lim3} \cite{Dav, Schiff}. The ratio of these spreads is:
\begin{equation}
\frac{\Delta E_{\text{Schr\"{o}dinger}}}{\Delta E_{\text{Sommerfeld}}} = \frac{4n}{2n - 1}  \qquad \left(n = 2, 3, \dots\right).
\end{equation}
When $n = 2$, one obtains ${\Delta E_{\text{Schr\"{o}dinger}}} = (8/3)\, {\Delta E_{\text{Sommerfeld}}}$.
\noindent\textbf{Note.}
With the help of Mathematica, we derived the next two terms in the expansion~\eqref{lim3} as follows \cite{BarleySusMath, SomAS}:%
\begin{equation}
-\frac{\mu ^{6}}{4n^{6}}\left[ \frac{5}{4}-\frac{3n}{j+1/2}+\frac{3n^{2}}{%
2\left( j+1/2\right) ^{2}}+\frac{n^{3}}{2\left( j+1/2\right) ^{3}}\right],%
\label{mu6}
\end{equation}
and
\begin{equation}
\frac{\mu ^{8}}{16n^{8}}\left[ \frac{35}{8}-\frac{15n}{j+1/2}+\frac{15n^{2}}{%
\left( j+1/2\right) ^{2}}-\frac{n^{3}}{\left( j+1/2\right) ^{3}}-\frac{3n^{4}%
}{\left( j+1/2\right) ^{4}}-\frac{n^{5}}{\left( j+1/2\right) ^{5}}\right]. 
\label{mu8}
\end{equation}

Hence, Sommerfeld's fine structure formula can now be derived semiclassically from the radial Dirac equations. This derivation not only recovers the quantization rules of Bohr and Sommerfeld \cite{SomAS, SomQM}—introduced nearly a decade before the concept of spin—%
\footnote{The concept of electron spin was introduced by G.~E.~Uhlenbeck and S.~Goudsmit in a letter published in {\it Die Naturwissenschaften}, issue of 20 November 1925; see \cite{Mehra} for details.
The spin concept explained why the famous Stern--Gerlach experiment \cite[pp.~124--126, 505--507]{SomAS} ends up with two separated beams of silver atoms, in contrast to the prediction of `old' quantum theory without spin, which would result in the prediction of three separated beams of silver atoms \cite{Stern}. 
}
but also links them to modern quantum theory.
Indeed, the classical relativistic Hamiltonian does not include spin, creating an inherent ambiguity in Bohr–Sommerfeld quantization (see \cite{GranUFN, PetrovUFN}). %
{\textit{De facto\/,}} we have completed Sommerfeld's original WKB arguments \cite[pp.~134--143]{SomQM} from a modern mathematical perspective.

For a full analytical treatment, including the non-relativistic limit, see Refs.~\cite{EEKS-SKS, Ni:Uv, Sus:Trey2008, Sus:Vega:Barl} 
(based on the Nikiforov--Uvarov method), as well as standard texts \cite{AkhBer, BerLifPit, Dav, Rose, Schiff}.

{\scshape Summary:} The principal result of Sommerfeld’s fine structure theory, namely, formulas~(\ref{SommDir})--(\ref{SommDirNu}), remains the correct expression for energy levels in hydrogen-like systems within wave mechanics.
By a remarkable twist of historical fate, Sommerfeld managed in 1916 to derive the correct formula from what ultimately proved to be an inadequate theoretical framework.
One must recall that, at the time, both quantum mechanics and the concept of spin were still nearly a decade away.
Thus, “perhaps the most remarkable numerical coincidence in the history of physics,” as Kronig remarked, illustrates the curious fact that flawed physical models can nevertheless yield correct quantitative predictions 
(see \cite[pp.~84--85 and references therein]{Kragh1985}).
%

%%%%% Figure 6, Sommerfeld's calculations %%%%%
%
\begin{figure}[hbt!]
\centering
\includegraphics[width=0.775\textwidth]{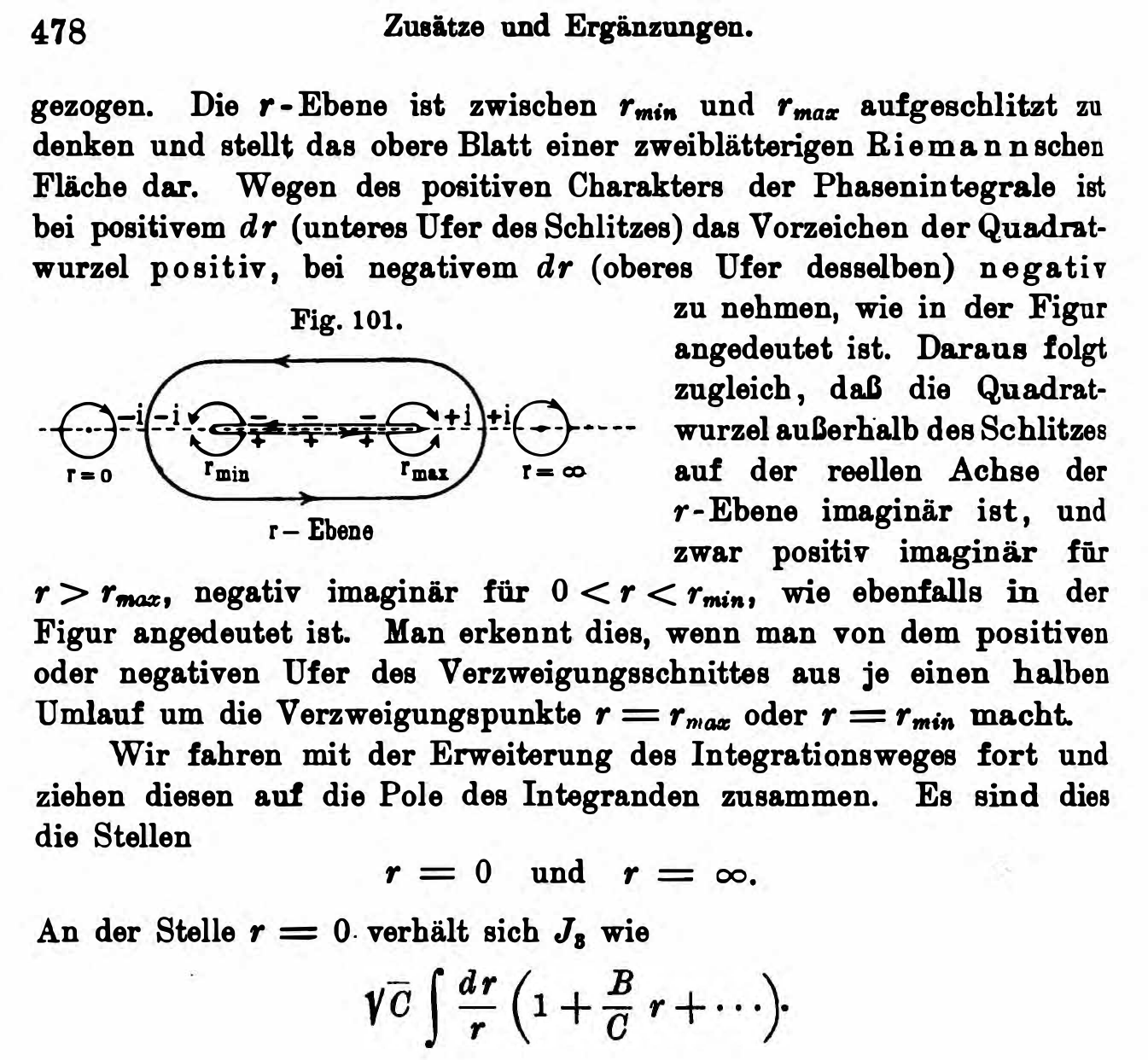} 
\caption{{A half-page from Sommerfeld's book showing the contour of
integration. {\small{\ 
\url{https://archive.org/details/atombauundspekt00sommgoog/page/478/mode/2up}% 
}}} (See \cite[pp.~468--470]{Goldstein} and \cite[pp.~611--612]{SomAS} for more details.)}
\label{Figure4}
\end{figure}

%
%%===========================================
%%% Section %%%
%%===========================================
%

\section{Evaluation of the Sommerfeld-Type Integrals}\label{sec:SommerfeldIntegral} 

Teaching mathematics and calculus in the United States has developed a “modern tradition” of favoring so-called “real world problems” — and this is indeed “The One” remarkable example:
the integral required for the semiclassical derivation of one of the most remarkable formulas of the last century — the Sommerfeld fine structure formula.

For all problems under consideration, we employ the generic integral (\ref{SomABC})--(\ref{SomABCInt}),
originally evaluated by Sommerfeld using complex analysis \cite[pp.~611--612]{SomAS} (see Figure~\ref{Figure4}).
In contrast, an elementary derivation of this integral was presented in \cite{Barleyetal2021}.
By integrating by parts on the left-hand side of (\ref{SomABCInt}), one obtains:
\begin{align}
& {\left. \int_{r_{1}}^{r_{2}}p(r)\,dr = r p(r) \right\vert }_{r_{1}}^{r_{2}} - 
\int_{r_{1}}^{r_{2}}\dfrac{r\left[ -(B/r^{2})+2(C/r^{3})\right] }{2p(r)}\,dr 
\label{B1} \\
& \quad = \dfrac{B}{2}\int_{r_{1}}^{r_{2}}\dfrac{dr}{\sqrt{-A r^{2} + B r - C}} 
- \int_{r_{1}}^{r_{2}}\dfrac{(C/r^{2})\,dr}{\sqrt{-A + (B/r) - (C/r^{2})}}. 
\notag
\end{align}

For the penultimate integral, we write:
\begin{eqnarray}
&&\int_{r_{1}}^{r_{2}}\dfrac{dr}{\sqrt{\left( \frac{B^{2}}{4A} - C \right)
- \left( r \sqrt{A} - \frac{B}{2\sqrt{A}} \right)^2 }}  \notag \\
&&\quad = {\left. \dfrac{1}{\sqrt{A}} \arcsin \dfrac{r\sqrt{A} - \frac{B}{2\sqrt{A}}}
{\sqrt{\frac{B^{2}}{4A} - C}} \right\vert }_{r_{1}}^{r_{2}} = \dfrac{\pi}{\sqrt{A}}. 
\label{B2}
\end{eqnarray}

Next, applying the substitution $r = 1/x$ in the final integral of (\ref{B1}), we obtain:
\begin{eqnarray}
&& -\int_{x_{1}=1/r_{1}}^{x_{2}=1/r_{2}}\dfrac{C\,dx}{\sqrt{-A + B x - C x^{2}}}
\notag \\
&=& -\sqrt{C} \int_{x_{1}}^{x_{2}} \dfrac{\sqrt{C}\,dx}{\sqrt{\left( \frac{B^{2}}{4C} - A \right) 
- \left( x \sqrt{C} - \frac{B}{2\sqrt{C}} \right)^2}} = \pi \sqrt{C}, 
\label{B3}
\end{eqnarray}
where we completed the square and evaluated a standard definite integral.
(Alternatively, one may interchange $A$ and $C$ and follow a similar route.)
Combining the results from the last two integrals completes the proof.

An additional independent evaluation of the Sommerfeld-type integral is presented in Appendix~\ref{appendix:c}.

%
%%===========================================
%%% Section %%%
%%===========================================
%

\section{Further Examples and Resolution of ``Sommerfeld's Puzzle"}

As is well known, Bohr introduced semiclassical quantization rules for
hydrogen-like atoms based on classical circular motion, while Sommerfeld
extended these ideas to relativistic elliptical orbits \cite{SomAS}. Measurements of
fine-structure splitting by Paschen were interpreted as experimental tests of
the special theory of relativity \cite{Biedenharn1983, Kragh1985, SomAS, Somm1940}.
The exact solution was obtained for the first time by C.~G.~Darwin \cite{Darwin1928} 
and W.~Gordon \cite{Gordon1928}, but only after the discovery of the Dirac equation \cite{Dirac1928, Dirac1977} — 
\textit{and remarkably, the new result reproduced the ``old" Sommerfeld formula}~(\ref{SommDir})!

Werner Heisenberg \cite{HeisenbergNobPrize} referred to this agreement as a `miracle' and wrote: %  
``{\textit{It would be intriguing to explore whether this is about a miracle or it is
the group-theoretical approach which leads to this formula}}''~\cite{Heisenberg}. 
In a 1956 letter, Erwin Schr\"{o}dinger commented: %
``{\textit{%
This is a fortuitous coincidence}}''~\cite{YouMan}. As demonstrated in \cite{SusPuzz}, 
Schr\"{o}dinger appears to have been correct — the ``Sommerfeld Puzzle" \cite{Biedenharn1983} 
has now been resolved and extended to a class of
multi-dimensional problems involving different symmetry groups.

{\scshape{Topic to review:}} The Nikiforov--Uvarov approach \cite[pp.~97--98]{Barleyetal2021}, \cite[pp.~44--47]{EEKS-SKS}, \cite[pp.~339--347]{NikNovUv}, and \cite{Ni:Uv}.

For exact solutions, the generalized equation of hypergeometric type \cite{Ni:Uv} is given by:
\begin{equation}
u^{\prime \prime }+\frac{\widetilde{\tau }(x)}{\sigma (x)}u^{\prime }+\frac{%
\widetilde{\sigma }(x)}{\sigma ^{2}(x)}u=0 , \label{NU1}
\end{equation}%
with the parameter choices:%
\begin{eqnarray}
\sigma (x) &=&x,\qquad \widetilde{\tau }(x)=0,  \label{STP1} \\
\widetilde{\sigma }(x) &=&-ax^{2}+bx-c+\frac{1}{4}.  \notag
\end{eqnarray}%
Then:%
\begin{eqnarray}
\pi (x) &=&\frac{\sigma ^{\prime }-\widetilde{\tau }}{2}\pm \sqrt{\left( 
\frac{\sigma ^{\prime }-\widetilde{\tau }}{2}\right) ^{2}-\widetilde{\sigma }%
+k\sigma }  \nonumber  \label{NUPi} \\
&=&\frac{1}{2}\pm \sqrt{ax^{2}+(k-b)x+c}
\end{eqnarray}
must reduce to a linear function \cite{Ni:Uv}. 
When $k=b\pm 2\sqrt{ac}$, one can complete the square to obtain:%
\begin{equation}
\pi =\frac{1}{2}\pm \left( \sqrt{a}x\pm \sqrt{c}\right) ,
\quad \tau =  \widetilde{\tau } +2\pi = 1 \pm 2 \left( \sqrt{a}x\pm \sqrt{c}\right).
\label{STP23}
\end{equation}%
We then choose:%
\begin{equation}
\tau ^{\prime }=-2\sqrt{a}<0\quad \text{and} \quad \lambda =k+\pi' =b-2\sqrt{ac}-%
\sqrt{a} . \label{STP4}
\end{equation}

As a result, for all Sommerfeld-type potentials, the Nikiforov--Uvarov
quantization rule \cite{Ni:Uv}:
\begin{equation}
\lambda +n\tau ^{\prime }+\frac{1}{2}n\left( n-1\right) \sigma ^{\prime
\prime }=0\qquad (n=0,1,2,\dots) ,  \label{NU7}
\end{equation}%
yields:%
\begin{equation}
\dfrac{b}{2\sqrt{a}}-\sqrt{c}=n+\dfrac{1}{2},  \label{STP5}
\end{equation}%
as a generic formula for the exact energy levels with $n=n_r$. 
(It is worth noting that Sommerfeld had already obtained a similar relation in special cases \cite{SomQM}.)

\smallskip
{\scshape{The puzzle resolution:}}

By comparing (\ref{wkbBoSomSus}) and (\ref{STP5}), we arrive at the following result \cite{SusPuzz}:
\medskip

\begin{theorem}
\begin{equation}
a=A,\qquad b=B,\qquad c=C.  \label{Resolution}
\end{equation}
\end{theorem}

\smallskip 
Indeed, the generic (WKB-based) rule~(\ref{wkbBoSomSus}) is also valid for the exact energy
levels (\ref{STP5}) obtained via the Nikiforov--Uvarov approach \cite{EEKS-SKS, SusPuzz}, for all Coulomb problems under consideration.
Other examples include quantum harmonic oscillators and systems with 
Kratzer and P\"{o}schl--Teller potentials \cite{EEKS-SKS}. (See also \cite{Kryuch:Sus:Vega12} for an extension 
of Schr\"{o}dinger's coherent states \cite{SchrodingerCohrent}.)

%\vfill\eject\newpage

In connection with Sommerfeld's fine-structure formula, Erwin Schr\"{o}dinger testified, {\it inter alia},
in a letter dated 29 February 1956 \cite{YouMan}:
{{\it{%
{\textquotedblleft ... you are naturally aware of the fact that Sommerfeld derivation of the fine-structure formula provides only fortuitously
the result demanded by the experiment.
One may notice then from this particular example that newer form of quantum theory $($i.~e., quantum mechanics\/$)$
is by no means such an inventible continuation of the older theory as is commonly supposed.
Admittedly the Schr\"{o}dinger theory, relativistically framed $($without spin\/$)$, gives a {\it{formal}} expression of the fine-structure formula
of Sommerfeld, but it is {\it{incorrect}} owing to the appearance of half-integers instead of integers.
My paper in which this is shown has ... never been published; it was withdrawn by me and replaced by non-relativistic treatment...
The computation $[$by the relativistic method\/$]$ is far too little known.
It shows in one respect how {\it{necessary}} Dirac's improvement was, and on the other hand it is wrong to assume that the older form of quantum theory is `broadly' in accordance with the newer form.%
\/\textquotedblright}}}}

\smallskip

{\scshape{Methodological note:}}  
It should now be clear that only after the “two quantum revolutions” could the ambiguity in the quantization of the Kepler problem in the `old quantum mechanics' \cite{GranUFN, PetrovUFN} be resolved—namely, when spherical symmetry is explicitly taken into account through separation of variables, and the corresponding radial equations are derived exactly, without approximation, and then subjected to the WKB method under Langer’s correction.

Schr\"{o}dinger appears to have been the first to follow this route in 1925
(or was close to doing so but avoided a misstep by switching to the exact solution?) in his unpublished notes. He used equation (\ref{wkb5.10}), which corresponds to a spin-zero particle (see \cite{Barleyetal2021}, Figure~\ref{Figure3}, for his original notebook, and \cite{Kryuchkovetal} for further discussion on the concept of spin and the wave equations). 
As later attested, due to the discrepancy with experimental results, Schr\"{o}dinger never published this work (see also \cite[Appendix~D]{Barleyetal2021}
for his letter to Weyl).

%%===========================================
%%% Section %%%
%%===========================================
%

\section{A Mistake that Schr\"{o}dinger Never Made} 

Interestingly, in Figure~\ref{Figure3}, the lower left-hand portion displays the `old' Bohr--Sommerfeld quantization rule
with what is now recognized as the Langer correction \cite{BerryMount-SKS},
namely,% 
\footnote{In this case, one obtains $\sqrt{C} = n + 1/2$.}
$n(n+1) \rightarrow \left(n+\frac{1}{2}\right)^2$, but
with $l$ rather than $l+\frac{1}{2}$ on the right-hand side, as it should appear in the WKB approximation (\ref{wkbBoSom}) 
(in Schr\"{o}dinger's notation, you may wish to interchange $n \leftrightarrow l$, as is customary today; see also footnotes$^{14\text{--}15}$ on p.~25 below).
As we now understand well, this form would yield an incorrect spectrum for his relativistic equation ---
in particular, the principal quantum number in the non-relativistic Kepler problem would acquire half-integer values rather than integers
(see our supplementary Mathematica notebook for further details \cite{BarleySusMath}).

However, Schr\"{o}dinger never made this mistake. Instead, he outlined, ``...({\it{although having used just written words}})..." \cite[Appendix~D, p.~100]{Barleyetal2021}; see also \cite[184\dag\;~pp.~484--485]{Meyenn2011}, 
the exact solution via the Laplace method, expressed as a contour integral \cite[Appendix~C]{Barleyetal2021}!

Schr\"{o}dinger knew the Sommerfeld-type integral intimately. For example, it is explicitly referenced in his letter to Sommerfeld dated January 29, 1926 \cite[p.~462]{MEHRAIII}; see also \cite[041\dag~pp.~170--172]{Meyenn2011}: 
``{\textit{%
... Finally, I still wish to add that the discovery of the whole connection
$[$between the wave equation and the quantization of the hydrogen atom$]$ goes back to your beautiful quantization method for evaluating the radial quantum integral. It was the characteristic 
$- \frac{B}{\sqrt{A}} + \sqrt{C'}$, 
which suddenly shone out from the exponents ${\alpha_1}$ and ${\alpha_2}$ like a Holy Grail.%
}}''

In this letter, Schr\"{o}dinger reported for the first time the success of the wave theory in solving the quantum oscillator, rotator, the non-relativistic (and partially relativistic) hydrogen atom (Kepler problems), and the free motion of a point mass in infinite space and in a box, prior to the formal publications \cite{SchrQMI, SchrQMII}.
He also formulated a program for future research. 
For the reader’s benefit, the complete letter has been translated from German to English in Appendix~\ref{appendix:d}.

In a letter dated February 3, 1926 \cite[042\dag~pp.~173--175]{Meyenn2011}, Sommerfeld responded enthusiastically: 
``{\textit{What you write, in your essay and letter, is terribly interesting. My personal opinion on the mysticism of integers must remain silent, as must my personal convenience ... My impression is this: Your method is a substitute for the new quantum mechanics of Heisenberg, Born, Dirac ...
Because your results are completely consistent with theirs...}}"
This marked the beginning of the triumph of Schr\"{o}dinger’s wave mechanics \cite[pp.~617--636]{MEHRAIII}.
{\scshape{Timetable:}}
The exact dates of Schr\"{o}dinger's foundational discoveries, leading to his first publications \cite{SchrQMI, SchrQMII}, are not precisely recorded \cite{Kragh1982, Kragh1984, Wessels1979} and \cite[pp.~459--465]{MEHRAIII}.
However, one can estimate the timeline based on his letter to Einstein \cite[030\dag\;~pp.~141--143]{Meyenn2011}, dated November 3, 1925; Bloch’s recollection of two colloquia in Z\"{u}rich \cite{Bloch1976}, presumably held in late November and/or early December 1925 \cite[pp.~419--423]{MEHRAIII}; a letter to Wien \cite[037\dag~pp.~162--165]{Meyenn2011} from Arosa on December 27, 1925; and a letter to Sommerfeld \cite[041\dag~pp.~170--172]{Meyenn2011} from Z\"{u}rich on January 29, 1926.
This yields a reasonable estimate spanning from early November 1925 to the end of January 1926.

At the same time, in January 1926, Bohr reflected on the development of the `old' theory in a letter to his friend, the Swedish physicist Carl Oseen \cite[p.~85]{Kragh1985} (quoted from \cite[p.~73]{NB}):
``{\textit{%
At the present stage of the development of quantum theory, we can
hardly say whether it was good or bad luck that the properties of the
Kepler motion could be brought into such simple connection with the
hydrogen spectrum, as was believed possible at one time.
If this connection had merely had that asymptotic character which one might
expect from the correspondence principle, then we should not have been
tempted to apply mechanics as crudely as we believed possible for some
time. On the other hand, it was just these mechanical considerations
that were helpful in building up the analysis of the optical phenomena
which gradually led to quantum mechanics.%
}}''
--- It was, indeed, hard to predict!

%\vfill\eject\newpage
%%===========================================
%%% Section %%%
%%===========================================
%

\section{Conclusion}

Traditional textbooks \cite{AkhBer, BerLifPit, Dav, La:Lif, MerzB, Schiff} do not address the derivation of the Sommerfeld fine structure formula 
via semiclassical approximation—and now the reader can appreciate why.
Indeed, {\it{de facto,}} there are three distinct levels of complexity, reflecting the historical ``Three Quantum Revolutions" in the development of quantum physics:
-- The elementary level of classical mechanics \cite{Gantmacher, Goldstein}, in the Bohr model of the atom;
-- Introductory quantum mechanics \cite{La:Lif, MerzB}, covering the non-relativistic (and unsuccessfully relativistic) hydrogen atom;
-- And finally, relativistic quantum theory \cite{AkhBer, BerLifPit, Dav, Rose, Schiff}, involving the Dirac equation and 
the Sommerfeld fine structure formula. 
\vfill\eject\newpage 
%

%\noindent
This history line is schematically presented in the following timetable:
%
%All of these are schematically represented in the following time-table:
%

%\newpage

\begin{tabular}{|l|l|}
\hline
Time Up & Main Results \\ \hline
\begin{tabular}{l}
Dirac's wave mechanics: \\ 
1928%
\end{tabular}
& 
\begin{tabular}{l}
Relativistic Dirac's equation; \\ 
Fine structure formula (\ref{SommDir})--(\ref{SommDirNu}); \\ 
we derive this result in the WKB approximation.%
\end{tabular}
\\ \hline
\begin{tabular}{l}
Schr\"{o}dinger's wave \\ 
 mechanics: 1925--1926%
\end{tabular}
& 
\begin{tabular}{l}
Non-relativistic and relativistic Schr\"{o}dinger's equations: \\ 
Bohr's formula (\ref{wkb5.9}), \\ 
Schr\"{o}dinger's fine structure formula (\ref{sol16})--(\ref{lim2}); \\ 
we derive (\ref{wkb5.9}) and (\ref{sol16})--(\ref{lim2}) in the WKB
approximation.%
\end{tabular}
\\ \hline
\begin{tabular}{l}
`Old' quantum mechanics: \\ 
1911--1916%
\end{tabular}
& 
\begin{tabular}{l}
Rutherford's planetary model of the atom; \\ 
Bohr--–(Wilson)--Sommerfeld quantization rules: \\ 
Bohr's formula (\ref{BM6}) and the fine structure formula (\ref%
{BidSomEnd})%
\end{tabular}
\\ \hline
\end{tabular}

%\newpage
\bigskip
%
%%%%% Figure Sommerfeld and Bohr  %%%%%%
\begin{figure}[hbt!]
\centering
\includegraphics[width=0.75\textwidth]{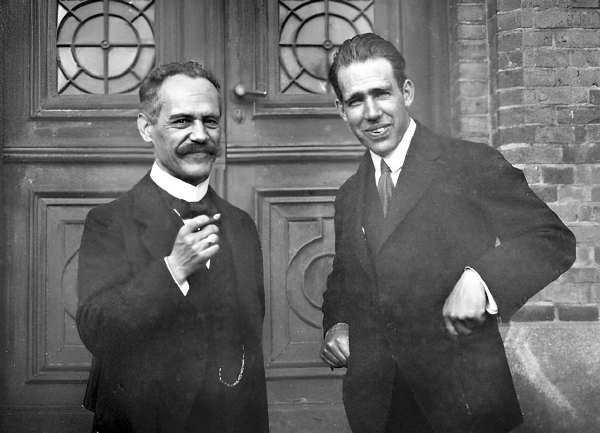}
%{sommerfeld-bohr.jpg}
\caption{Arnold Sommerfeld and Niels Bohr in Lund, September~1919, 
during his impressive lecture tour of Scandinavia after World War~I:
{\textit{... a sudden return to the good times before the war\/}}
\cite[pp.~240--246]{Eckert}. 
{%\textit
{Courtesy of Deutsches Museum, Munich\/.}}}
\label{Bohr-Sommerfeld}
\end{figure}
%

%\medskip

The WKB method becomes applicable only after a careful study of the basic properties of the Dirac equation, including the construction of spinor spherical harmonics \cite{AkhBer, BerLifPit, Rose, Sus:Trey2008, Sus:Vega:Barl, Varshalovich1988}, and a nontrivial separation of variables in spherical coordinates. 
Naturally, this poses a ``pedagogical challenge" and requires significant time in both teaching and learning quantum mechanics.
These notes—originally motivated by introductory courses on the mathematics of quantum mechanics at Arizona State University {\cite{Barleyetal2021, EEKS-SKS, GorBarSus23, Kryuchkovetal, Koutetal, Kryuch:Sus:Vega12, SusPuzz, Sus:Trey2008, Sus:Vega:Barl}} 
and at Technische Universit\"{a}t M\"{u}nchen \cite{Ruffing}—though not exhaustive, may help readers bridge this gap. 
%
%\medskip

The resolution of the ``Sommerfeld Puzzle" remains of theoretical and pedagogical interest \cite{SusPuzz}.
Only in Dirac's theory, which gave a fully relativistic quantum-mechanical description of the hydrogen atom 
and accurately predicted its fine structure, can this question be formulated unambiguously. ---
And only at this level can we arrive not at a “puzzle”, but at a mathematically well-defined statement,
such as a hypothesis or a theorem.
%
%\medskip

%
Computer algebra systems such as Maple and Mathematica are valuable tools for teaching and learning quantum mechanics, particularly for beginners.
These systems can handle complex mathematical calculations, allowing students to focus on understanding the underlying concepts and problem-solving strategies rather than getting bogged down in tedious computations. 
One should admit nonetheless that each of the quantum mechanical problems under consideration, with exception of a few trivial ones, usually requires a separate analysis with lots of specific details that are far away from a formal application of a given computer algebra system, as it might be thought of at the first glance (see, for example, \cite{EEKS-SKS, GorBarSus23, Koutetal}).

Ultimately, the practical use of Mathematica in this article \cite{BarleySusMath} can also aid students 
in tackling complex computations and better understanding quantum physics.
Artificial Intelligence calculation with its linear functions and simple activation functions clearly cannot, even remotely, achieve the level of correct mathematical derivation of, say, equations (\ref{BidSomEnd})--(\ref{BidSom20}) provided in Appendix~\ref{appendix:e}. % 
%
%Although the application of computer algebra to quantum physics can nowadays be enhanced by an artificial intelligence, hard work with %Mathematica codes remains the main criterion for success.

\smallskip
 
\noindent \textbf{Acknowledgments} We are grateful to Dr.~Ruben~Abagyan, Dr.~Sergey~I.~Kryuchkov, Dr.~Nathan A.~Lanfear, 
and Dr.~Eugene Stepanov for their assistance and insightful comments.

\newpage
%

%\vfill\eject\newpage 
%%===========================================
%%% APPENDIX A %%%
%%===========================================
%
\appendix

\section{Velocity, Acceleration and Angular Momentum for \newline
the Uniform Circular Motion}\label{appendix:a}

For a uniform circular motion one gets%
\begin{equation}
a=\frac{v^{2}}{r}.  \label{A1}
\end{equation}%
%
%
%%% Figure 7, a uniform circular motion  %%%%
%
\begin{figure}[hbt!]
\centering
\includegraphics[width=0.415\textwidth]{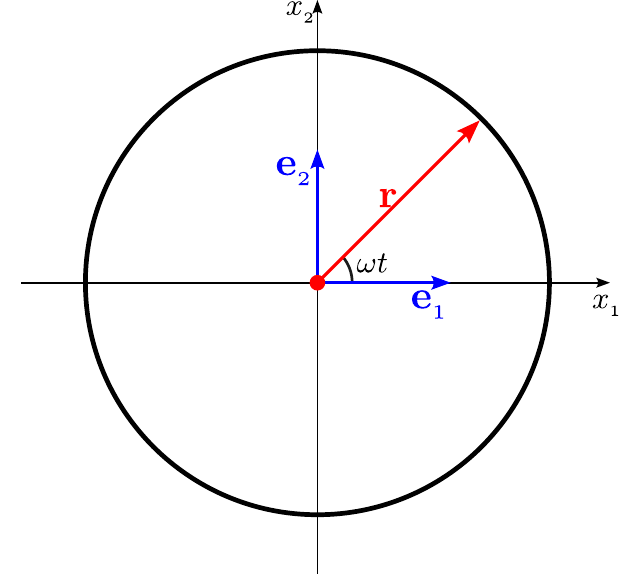}
\caption{Uniform circular motion.}
\label{Figure5}
\end{figure}

Indeed, if%
\begin{equation}
\mathbf{r}=\mathbf{r}\left( t\right) =r\cos \left( \omega t\right) \mathbf{e}%
_{1}+r\sin \left( \omega t\right) \mathbf{e}_{2},\qquad \mathbf{r}\left(
0\right) =r\mathbf{e}_{1},  \label{A2}
\end{equation}%
where $\mathbf{e}_{1}$ and $\mathbf{e}_{2}$ are two orthonormal vectors
(Figure~\ref{Figure5}), then%
\begin{equation}
\mathbf{v}=\frac{d\mathbf{r}}{dt}=r\omega \left( -\sin \left( \omega
t\right) \mathbf{e}_{1}+\cos \left( \omega t\right) \mathbf{e}_{2}\right)
,\qquad \mathbf{r}\cdot \mathbf{v}=0,  \label{A3}
\end{equation}%
and%
\begin{equation}
\mathbf{a}=\frac{d\mathbf{v}}{dt}=-r\omega ^{2}\left( \cos \left( \omega
t\right) \mathbf{e}_{1}+\sin \left( \omega t\right) \mathbf{e}_{2}\right)
=-\omega ^{2}\mathbf{r}.  \label{A4}
\end{equation}%
Thus,%
\begin{equation}
\mathbf{v}^{2}=\mathbf{v}\cdot \mathbf{v}=r^{2}\omega ^{2}\left( \sin
^{2}\left( \omega t\right) +\cos ^{2}\left( \omega t\right) \right)
=r^{2}\omega ^{2}=v^{2},\quad v=\omega r.  \label{A5}
\end{equation}%
In a similar fashion,%
\begin{equation}
\mathbf{a}^{2}=\mathbf{a}\cdot \mathbf{a}=\omega ^{4}r^{2},\qquad a=\omega
^{2}r.  \label{A6}
\end{equation}%
Relation (\ref{A1}) follows from the last two expressions: $a=\omega ^{2}r$
and $\omega =v/r.$

Moreover,%
\begin{eqnarray}
\mathbf{r}\times \mathbf{v} &=&r^{2}\omega \left\vert 
\begin{array}{ccc}
\mathbf{e}_{1} & \mathbf{e}_{2} & \mathbf{e}_{3} \\ 
\cos \left( \omega t\right) & \sin \left( \omega t\right) & 0 \\ 
-\sin \left( \omega t\right) & \cos \left( \omega t\right) & 0%
\end{array}%
\right\vert  \label{A7} \\
&=&r^{2}\omega \left( \cos ^{2}\left( \omega t\right) +\sin ^{2}\left(
\omega t\right) \right) \mathbf{e}_{3}=rv\mathbf{e}_{3}  \notag
\end{eqnarray}%
(see Figure~\ref{Figure1}). Eq.~(\ref{BidSom2}) can be derived in a similar way.

\newpage
%%===========================================
%%% APPENDIX B %%%
%%===========================================

\section{Instability of a Hydrogen Atom in Classical Physics}\label{appendix:b}

As is known \cite{BohrAtom, MilUFN, Reed, Stein}, a rotating electron in Rutherford's planetary
model must fall into the nucleus---a spiral-in---according to the laws of classical mechanics
and electrodynamics. Indeed, the total instantaneous power emitted over all solid angles
is given by the well-known result of Larmor \cite{Jackson}: 
\begin{equation}
\dfrac{dE}{dt} = -\dfrac{2e^{2}a^{2}}{3c^{3}},  \label{C1}
\end{equation}
where \( c \approx 2.\,997\,9 \times 10^{10} \) cm/s is the speed of light in cgs
units. Here, according to (\ref{BM1})--(\ref{BM3}), 
\begin{equation}
a = \dfrac{v^{2}}{r} = \dfrac{e^{2}}{mr^{2}},  \label{C2}
\end{equation}
and thus,
\begin{equation}
\dfrac{dE}{dt} = -\dfrac{2e^{2}}{3c^{3}} \left( \dfrac{e^{2}}{mr^{2}} \right)^{2} 
= -\dfrac{2e^{6}}{3c^{3}m^{2}r^{4}}.  \label{C3}
\end{equation}
On the other hand, from the virial theorem (\ref{BM3}), one gets: 
\begin{equation}
\dfrac{dE}{dt} = \dfrac{e^{2}}{2r^{2}} \dfrac{dr}{dt}.  \label{C4}
\end{equation}
Equating (\ref{C3}) and (\ref{C4}), we obtain:
\begin{equation}
r^{2} \dfrac{dr}{dt} = -\dfrac{4e^{4}}{3m^{2}c^{3}}, \qquad \text{or\quad}
3r^{2}\,dr = -\dfrac{4e^{4}}{m^{2}c^{3}}\,dt.  \label{C5}
\end{equation}
Integrating both sides yields:
\begin{equation}
-r_{1}^{3} = \left. r^{3} \right\vert_{r_{1}}^{0} = \int_{r=r_{1}}^{0} 3r^{2}\,dr 
= -\dfrac{4e^{4}}{m^{2}c^{3}} \int_{t=0}^{\tau} dt 
= -\dfrac{4e^{4}}{m^{2}c^{3}}\,\tau,  \label{C6}
\end{equation}
where \( r_{1} \approx .\,529\,21 \times 10^{-8} \) cm is the first Bohr radius of
a hydrogen atom, as given in (\ref{BM5}). Therefore, an electron in Rutherford's model would fall into
the nucleus in less than a nanosecond:
\begin{equation}
\tau = \dfrac{m^{2}c^{3}}{4e^{4}}\,r_{1}^{3} \approx 
\dfrac{(9.\,109\,4 \times 10^{-28})^{2} (2.\,997\,9 \times 10^{10})^{3}}{
4 (4.\,803\,2 \times 10^{-10})^{4}} \, (.\,529\,21 \times 10^{-8})^{3}
\approx 1.\,556\,4 \times 10^{-11}\ \mathrm{s}  \label{C7}
\end{equation}
(according to \cite{MilUFN}, this estimate was obtained in 1904 by G.~A.~Schott).

The electron velocity on the first Bohr orbit can be estimated as follows:
\begin{equation}
v_{1}=\frac{e}{\sqrt{mr_{1}}}\approx \frac{4.\,803\,2\times 10^{-10}}{\left(%
.\,529\,21\times 10^{-8}\cdot  9.\,109\,4\times 10^{-28}\right)^{1/2}}\approx 2.\,
187\,6\times 10^{8} \ \text{cm/s\ }\approx .7\,297\,\times 10^{-2}\ c\label{C8}
\end{equation}%
(a non-relativistic motion), and for the time of one revolution we arrive at:
\begin{equation}
t_{\text{rot}}=\frac{2\pi r_{1}}{v_{1}}\approx 2\cdot 3.\,141\,5\ \frac{%
.\,529\,21\times 10^{-8}}{2.\,187\,6\times 10^{8}}\approx
1.\,519\,9\times 10^{-16}\ \mathrm{s}.\label{C9}
\end{equation}
Therefore, the total number of rotations before falling into the center can be
estimated as:
\begin{equation}
N_{{\rm{total}}}=\frac{\tau }{t_{\text{rot}}}\approx \frac{1.\,556\,4\times 10^{-11}}{%
1.\,519\,9\times 10^{-16}}\approx 102\,400.\label{C10}
\end{equation}%
The spiral-in time, \( \tau \), is much longer than the orbital time, \( t_{\text{rot}} \), so treating the spiral-in 
as a succession of circular orbits in a hydrogen atom is plausible (see \cite{Reed, Stein} for more details).

\newpage
%%===========================================
%%% APPENDIX C %%%
%%===========================================
\section{An Independent Evaluation of the Sommerfeld-type
Integrals}\label{appendix:c}

On the contrary, one can use the technique of differentiation with respect to
parameters for the familiar integrals related to the Bohr--Sommerfeld
quantization rule \cite{SomAS, SomQM}. As is well known, if%
\begin{equation}
J(x) = \int_{f(x)}^{g(x)} F(x,y)\ dy,  \label{J1}
\end{equation}
then%
\begin{equation}
\frac{dJ}{dx} = \int_{f(x)}^{g(x)} \frac{\partial F(x,y)}{\partial x}\ dy 
+ F(x, g(x))\,\frac{dg}{dx} - F(x, f(x))\,\frac{df}{dx}.  \label{J2}
\end{equation}
In the WKB case, the last two terms vanish because the limits are turning points where the integrand vanishes \cite{Ghatetal}.
%\smallskip
%

We now apply this procedure for an independent evaluation of the
\textquotedblleft Sommerfeld-type\textquotedblright\ integrals discussed
in this note \cite{SusPuzz}. Indeed,
\begin{equation}
I = \int_{r_{1}}^{r_{2}} p(r)\ dr, \qquad 
p(r) = \sqrt{ -A + \frac{B}{r} - \frac{C}{r^{2}} } 
\quad (A, C > 0),  \label{Som1}
\end{equation}
provided \( p(r_{1}) = p(r_{2}) = 0 \), one finds:
\begin{eqnarray}
\frac{dI}{dB} &=& \frac{1}{2} \int_{r_{1}}^{r_{2}} \frac{dr}{\sqrt{-Ar^{2} + Br - C}}  \label{Som2} \\
&=& \frac{1}{2\sqrt{A}} \int_{r_{1}}^{r_{2}} \frac{dr}{\sqrt{ \dfrac{B^{2} - 4AC}{4A^{2}} - \left(r - \dfrac{B}{2A}\right)^{2} }}  \notag \\
&=& \frac{1}{2\sqrt{A}} \left. \arcsin \left( \frac{2Ar - B}{\sqrt{B^{2} - 4AC}} \right) \right\vert_{r_{1}}^{r_{2}} 
= \frac{\pi}{2\sqrt{A}}.  \notag
\end{eqnarray}
As a result,
\begin{equation}
\frac{dI}{dB} = \frac{\pi}{2\sqrt{A}}, \qquad 
I(B_{0} = 2\sqrt{AC}) = 0,  \label{Som3}
\end{equation}
and by integration,
\begin{equation}
I = \pi \left( \dfrac{B}{2\sqrt{A}} - \sqrt{C} \right).  \label{Som4}
\end{equation}

It is perhaps the most simple way of this integral evaluation.
\vfill\eject\newpage
%
%%===========================================
%%% APPENDIX D %%%
%%===========================================
%

\bigskip

\section{A Letter from Schr{\"o}dinger to Sommerfeld} \label{appendix:d}
{\noindent {\textbf{Schr{\"o}dinger to Sommerfeld}} (translated from \cite[ 041\dag\;~pp.~170--172]{Meyenn2011})\\} 
%[%\textbf
[{\scshape{Author's Note}:} Semantic Similar Translation as one would use in Modern German]\\
%

%\vspace{0.125cm}
%
\bigskip
${}$\hfill Zurich, January 29, 1926%
\footnote{This letter is also printed in Arnold Sommerfeld, \textsl{Scientific Correspondence,} Vol.~2, pp.~236--238 [In German]:
Arnold Sommerfeld, {\textsl{Wissenschaftlicher Briefwechsel,}} Band~2, S.~236--238
.}
%
%%%%%%% Kamal's original %%%%%%%
%
\medskip

%\vfill\eject\newpage
\noindent
Most honorable Herr Professor,

\noindent
for a long time I didn't let hear anything from me,%
%It has been so long since I have not let you hear from me,%
%
{\footnote{Schr\"{o}dinger's last (extant) letter \cite[ 025\dag\;~pp.~132--135]{Meyenn2011} to Sommerfeld was written in July 1925.}}
and hence I am writing now to you in a quick manner, namely to avoid that you are going to write to me earlier -- on the following behalf:
you may have looked already to my quantum article which I have sent to Mr. Geheimrat Wien%
{\footnote{This was Schr\"odinger's first communication on wave mechanics \cite{SchrQMI}, which he had sent a few days earlier to one of the editors of \textit{Annalen der Physik,} Wilhelm Wien (see also Wien's reply \cite[ 043\dag\;~p.~177]{Meyenn2011}).}}%
%66
%
, for the annals, having kindly asked him to show it to you, before publishing it in the journal.
Of course, I am most curious to hear no other opinion -- first -- but yours, namely: whether
you are sharing the ambitious hope I have, which I am coupling to the derivation of quantum
equations, being derived from a Hamiltonian principle.%
\footnote{Cf.~\cite{Gantmacher, Goldstein}  -- See also Sommerfeld's reply \cite[ 042\dag\;~pp.~173--175]{Meyenn2011}.}
%67
%

So far, I have transferred some more mechanical problems into the new perception's world. As
far as my mathematics is sufficient, all evolves in a most beautiful manner -- it is no primitive
copy of old quantization rules, but differs from them at some characteristic points.%
{\footnote{According to Schr\"{o}dinger's former colleague Peter Paul Ewald \cite[p.~385]{Ewald}, Courant and Hilbert's seminal 1924 work on methods of mathematical physics \cite{Courant1924} ``enabled physicists to grasp the spirit of a unified mathematical method in an especially important field, marked by keywords: eigenvalues, eigenfunctions. These problems... arise in physics through oscillations of all kinds."
It is almost certain that this book was not available to Schr\"{o}dinger in Arosa \cite{Barleyetal2021}.
Only in the second article on wave mechanics does Schr\"{o}dinger thank his assistant E.~Fues for pointing out a
connection with the Hermite polynomials for the harmonic oscillator problem and acknowledge the relation of his wave
function in the ``Kepler problem" with the ``polynomials of Laguerre" \cite{SchrQMII}.
}}
%68

The harmonic oscillator is to be treated with the same analytic aids like the Kepler problem$^*$
(in the equation of vibration which one obtains to determine the function $\psi$, one has to introduce
the square of the abscissa as an independent variable).%
{\footnote{The ``Planck oscillator" is treated in Schr\"{o}dinger's second paper \cite{SchrQMII} as the first example.}}
%69
%
Again, there appears the remarkable case that an equation, which is not integrable by ordinary
quadratures, becomes precisely integrable in case of the eigenvalues, namely by elementary
functions -- this shows how nature, in some kind of lovely way, is interested in making its exploration
easier for us.
As for the eigenvalues (energy levels), there are:
%
%The eigenvalues (energy levels) are: % 
$\frac{2n+1}{2} \ h\nu,$
i.~e. the so-called half-integer quantization.%
\footnote{Cf.~\cite{SchrQMII}. Half-integer quantum numbers had also been introduced in Heisenberg's early attempts to describe the anomalous Zeeman effect.} 
%70
%
Although the quantum differences are unchanged here, I see a great meaning behind
it, since
$
\frac{2n+1}{2} %\ h \nu
$ 
is the arithmetic mean of $n$ and $n + 1.$

The rotator (dumbbell) in three dimensions (i.~e. with two variables
$\vartheta, \varphi $) is quite simple, the
eigenfunctions are ordinary spherical harmonics, the eigenvalues (energy levels) are
$ n(n+1) \frac{h^2}{8\pi^2 J} .$
This characteristic expression $n(n + 1)$ stems from the differential equation of the spherical
harmonics.
Again, I find it most delighting, not because of the considered case, but since we may wish
that for future investigations, that one might obtain $n(n+1)$ instead of $n^2,$
{\footnote{In the introduction to his supplement on wave mechanics \cite[p.~2]{SomQM}, Sommerfeld pointed out that these difficulties of the old quantum theory had now found their natural explanation through the new wave mechanics.}}
%71
%
where necessary
(regarding your formulas on intensity and the formulas for anomal Zeeman splitting).%
\footnote{As Sommerfeld noted in \cite[pp.~333, 476]{SomAS}, the appearance of $j(j+1)$ instead of $j^2$ in the Land\'{e} $g$-factor ``suggests that not one state $j$ but two adjacent quantum states $j$ and $j+1$ are physically relevant."}
%72
%
At the contrary, in case of the dumbbell, one has first to evaluate whether the explanation of
the band spectra will be damaged. But I don't think so. The effect is, as one easily recognizes,
given by a very slight difference between the linear term of the positive and the negative branch.
And this difference, if I remember right, indeed appears (or anyhow at least in a similar way --
I calculated all that yesterday and did not have the chance to compare with the experimental
values in detail.)

For the free motion of the mass point one receives that every energy value may appear if the
mass point is located in the infinite space. If it is in a box, which one has to consider as a
boundary condition for the $\psi$ function, one will obtain the same energy values as for the quantization
of the zig zag motion.
The eigenfunctions correspond in case of the free mass point -- presumed one calculates in a
relativistic way -- to the phase waves of de Broglie.%
\footnote{As we learn from his letter to Einstein \cite[030\dag\;~pp.~141--143]{Meyenn2011}, Schr\"{o}dinger was inspired to this work by the ``brilliant theses" (1925) of Louis de Broglie on the phase waves of electrons, which he had first encountered in early November 1925.}
%73
%
For the mass point in the box, these are standing eigenvibrations of the box volume being related to the dispersion law of the de Broglie
phase waves.
%
%for the mass point in the box, they are standing eigen-oscillations of the box volume with the dispersion law of de Broglie phase waves.

The next important task appears to be -- apart from the calculation of important special cases,
like Stark effect, Zeeman effect and relativistic Kepler motion -- the formulation of a rule for
intensity and polarization which has to replace the correspondence principle.%
{\footnote{See also the remarks in his letters to Lorentz \cite[055\dag\;~pp.~203--205]{Meyenn2011} 
and Wentzel \cite[068\dag\;~pp.~226--228]{Meyenn2011}.}} 
%74
%
Since I believe that the $\psi$ function really describes all those processes in the atom which are
the reason for light emission, the function has to give insight into these processes.
One has to investigate the intensity beats between two simultaneously excited eigenvibrations
and their (i.~e. the beats') spatial distribution. On this way, the rich mathematical theory (orthogonality
of eigenfunctions, meaning of the eigenvalues as extremal values of the Hamiltonian
integral etc.) will surely lead to simple results.

I am worried about the relativistic Kepler problem.%
{\footnote{Schr\"{o}dinger, as shown in his notes (Figure~\ref{Figure3}), had first proposed a relativistic wave equation before arriving at the well-known non-relativistic one \cite{Barleyetal2021}. See also his letters to Sommerfeld \cite[044\dag\;~pp.~178--184]{Meyenn2011} and Lorentz \cite[076\dag\;~pp.~252--261]{Meyenn2011}; together with historical studies \cite{Kragh1982, Kragh1984, MEHRAIII}.}}
%75
%
I am not sure whether it is true -- what I claim in the manuscript -- that the nuclear co-motion is that essential for the new method
of description. I would even rather withdraw my remark that this would already be the case
in the old method of description, and I would kindly ask you to drop this remark if you think
that is nonsense.%
{\footnote{Sommerfeld addressed this point in his next letter \cite[042\dag\;~pp.~173--175]{Meyenn2011}.}} 
%76

But as regards my perception, it is only the nuclear's motion which will provide help, otherwise
one would obtain half partial quanta – in contrast to the experience. And these half partial
quanta stem exactly from those $n(n + 1)$ of the spherical harmonics, what is on the other hand
so commendable.

Finally, I would like to say that the discovery of the whole connection of the structures goes
back -- even if it is not easy to recognize from outside -- to your beautiful integration method
for evaluating the radial quantum integral.
It was the characteristic and familiar
$- \frac{B}{\sqrt{A}} + \sqrt{C'}, $ 
which suddenly, like a Holy Grail, was shining from the exponents  $\alpha_1$ and $\alpha_2$.%
\footnote{Cf.~\cite[p.~611--612]{SomAS}.}
%77

I hope that you, most honorable Herr Professor, and all yours are doing well. \\
%I hope that you, most esteemed Professor, and all your family are well.
%

\noindent
With the best and most sincere greetings from house to house, 
%

%\vfill\eject\newpage
\noindent I always remain 

%\noindent
${}$\hfill 
Your faithful and grateful sincere $\qquad\qquad\qquad\qquad$
$\qquad\qquad\qquad\qquad$ 
%$\qquad\qquad\qquad\qquad$ 
%
\bigskip
%Yours faithfully, 

${}$\hfill 
E.~Schr{\"o}dinger

\medskip

P.~S. 
%Postscript: 
Innsbruck has not yet been officially decided.%
\footnote{Schr\"{o}dinger had also received an offer from Innsbruck (cf. also his letter to Sommerfeld \cite[025\dag\;~pp.~132--135]{Meyenn2011}~), which he declined in mid-March (cf. his letter \cite[052\dag\;~pp.~197--200]{Meyenn2011} 
of March~17, 1926, to Thirring).}
%78
%
But I think I'll stay here.
It is mainly Schweidler's departure to Vienna%
{\footnote{The previous Innsbruck chair, Egon von Schweidler (1873--1948), had just been appointed to the I. Physics Institute in Vienna. His successor in Innsbruck, after the chair was downgraded to a lectureship (cf.~\cite[044\dag\;~pp.~178--184]{Meyenn2011}), became Schr\"{o}dinger's friend Arthur March (1891--1957).}}
%79
% 
that decides for me.
Herzfeld wrote to me after speaking with you (I don't know if it was directly your opinion) that we should try to support Smekal.%
{\footnote{Adolf Smekal (1895--1959) did not receive an extraordinary professorship in Vienna until 1927, and in 1928 he accepted a call to Halle. Herzfeld was at that time preparing to accept a professorship at Johns Hopkins University in Baltimore.}}
%80
%
I think it will be difficult because March is named {\it{que es loco}} and has been supplying for quite some time.
But Thirring wants to do it anyway and wrote to me at the time, asking me to inform him after an official decision has been made.
It would certainly be more appropriate from a factual perspective.

Please don't assume my decision is final during the discussion. It would be unpleasant for both ministries. And on the other hand, I quite appreciate the delay because, after much effort, I'll finally get a little something, namely a new blackboard in the lecture hall and, hopefully, a little more funding for the seminar library.
%

%\medskip
\newpage

%%===========================================
%%% APPENDIX E %%%
%%===========================================

\section{Mathematica Derivation of Equations (\ref{BidSomEnd})--(\ref{BidSom20})}\label{appendix:e}

From our complementary Mathematica notebook \cite{BarleySusMath}:

\noindent
The reader may copy all Mathematica {\textit{In[$\#$]:=}} sections below into a new notebook and execute them one by one in order to repeat our calculations.

\begin{mat}

Introduce function $s(\theta) = \dfrac{1}{r}$ from $(\ref{BidSom9})$:

In[1]:= s[\[Theta]_] := (1 + \[Epsilon]*Cos[\[Omega]*\[Theta]])/(a*(1 - \[Epsilon]^2))

In[2]:= s[\[Theta]]
Out[2]= $\displaystyle \frac{1 + \epsilon \cos(\theta \omega)}{a (1 - \epsilon^2)}$

and evaluate its derivative:

In[3]:= D[s[\[Theta]], \[Theta]]
Out[3]= $\displaystyle -\frac{\epsilon \omega \sin(\theta \omega)}{a (1 - \epsilon^2)}$

The r.h.s. of $(\ref{BidSom4})$, evaluate:

In[4]:= (D[s[\[Theta]], \[Theta]])^2 + (s[\[Theta]])^2
Out[4]= $\displaystyle \frac{(1 + \epsilon \cos(\theta \omega))^2 + \epsilon^2 \omega^2 \sin^2(\theta \omega)}{a^2 (1 - \epsilon^2)^2}$

and substitute:

In[5]:= % /. Sin[\[Theta] \[Omega]]^2 -> 1 - (Cos[\[Theta] \[Omega]])^2
Out[5]= $\displaystyle \frac{(1 + \epsilon \cos(\theta \omega))^2 + \epsilon^2 \omega^2 (1 - \cos^2(\theta \omega))}{a^2 (1 - \epsilon^2)^2}$

Introduce $S$ and eliminate cosine:

In[6]:= Solve[S == (1 + \[Epsilon] Cos[\[Theta] \[Omega]])/(a (1 - \[Epsilon]^2)), Cos[\[Theta] \[Omega]]]  
Out[6]= $\displaystyle \left\{ \cos(\theta \omega) \to \frac{-1 + a S - a S \epsilon^2}{\epsilon} \right\}$

In[7]:= FullSimplify[%% /. Cos[\[Theta] \[Omega]] -> (-1 + a S - a S \[Epsilon]^2)/\[Epsilon]]  
Out[7]= $\displaystyle \frac{\omega^2 + a S \left(-2 \omega^2 - a S (-1 + \epsilon^2)(-1 + \omega^2)\right)}{a^2 (1 - \epsilon^2)}$


As a result, in terms of $S$, the above expression becomes:

In[8]:= Collect[%, S]  
Out[8]= $\displaystyle \frac{\omega^2}{a^2 (1 - \epsilon^2)} - \frac{2 S \omega^2}{a (1 - \epsilon^2)} - S^2 (-1 + \omega^2)$

Equation $(\ref{BidSom4})$, define the difference between r.h.s. and l.h.s.:

In[9]:= P = (En/(m c^2) + (Z (e^2) S )/(m c^2))^2 - 
  1 - ((\[HBar] Subscript[n, \[Theta]])/(
    m c))^2*(\[Omega]^2/(a^2 (-1 + \[Epsilon]^2)) - (2 S \[Omega]^2)/(
     a (-1 + \[Epsilon]^2)) - S^2 (-1 + \[Omega]^2))
Out[9]= $\displaystyle -1 + \left(\frac{En}{c^2 m} + \frac{e^2 S Z}{c^2 m}\right)^2 - \left(\frac{\omega^2}{a^2 (1 - \epsilon^2)} - \frac{2 S \omega^2}{a (1 - \epsilon^2)} - S^2 (-1 + \omega^2)\right) \frac{\hbar^2 n_\theta^2}{c^2 m^2}$

and simplify:

In[10]:= % /. e^2 -> \[Alpha]*\[HBar]*c ;

In[11]:= Collect[%, S]  
Out[11]= $\displaystyle 
-1 + \frac{En^2}{c^4 m^2} 
- \frac{\omega^2 \hbar^2 n_\theta^2}{a^2 c^2 m^2 (1 - \epsilon^2)} 
+ S \left( \frac{2 En Z \alpha \hbar}{c^3 m^2} + \frac{2 \omega^2 \hbar^2 n_\theta^2}{a c^2 m^2 (1 - \epsilon^2)} \right) 
+ S^2 \left( \frac{Z^2 \alpha^2 \hbar^2}{c^2 m^2} + \frac{(-1 + \omega^2) \hbar^2 n_\theta^2}{c^2 m^2} \right)$

To satisfy $(\ref{BidSom4})$, all coefficients of this quadratic polynomials must be equal to zero:

In[12]:= {Subscript[C, 
  0] = (Z^2 \[Alpha]^2 \[HBar]^2)/(c^2 m^2) + ((-1 + \[Omega]^2) \[HBar]^2 
\!\(\*SubsuperscriptBox[\(n\), \(\[Theta]\), \(2\)]\))/(c^2 m^2), 
 Subscript[C, 
  1] = (2 En Z \[Alpha] \[HBar])/(c^3 m^2) + (2 \[Omega]^2 \[HBar]^2 
\!\(\*SubsuperscriptBox[\(n\), \(\[Theta]\), \(2\)]\))/(
   a c^2 m^2 (-1 + \[Epsilon]^2)), 
 Subscript[C, 2] = -1 + En^2/(c^4 m^2) - (\[Omega]^2 \[HBar]^2 
\!\(\*SubsuperscriptBox[\(n\), \(\[Theta]\), \(2\)]\))/(
   a^2 c^2 m^2 (-1 + \[Epsilon]^2))}
Out[12]= $\left\{
\frac{Z^2 \alpha^2 \hbar^2}{c^2 m^2} + \frac{(-1 + \omega^2) \hbar^2 n_\theta^2}{c^2 m^2},\quad
\frac{2 En Z \alpha \hbar}{c^3 m^2} + \frac{2 \omega^2 \hbar^2 n_\theta^2}{a c^2 m^2 (1 - \epsilon^2)},\quad
-1 + \frac{En^2}{c^4 m^2} - \frac{\omega^2 \hbar^2 n_\theta^2}{a^2 c^2 m^2 (1 - \epsilon^2)}
\right\}$

From the vanishing leading term, we choose the positive solution:

In[13]:= Solve[Subscript[C, 0] == 0, \[Omega]]  
Out[13]= $\left\{
\omega \to -\frac{\sqrt{-Z^2 \alpha^2 + n_\theta^2}}{n_\theta},\quad
\omega \to \frac{\sqrt{-Z^2 \alpha^2 + n_\theta^2}}{n_\theta}
\right\}$


From the linear term, find the energy in terms of $a$ as follows:

In[14]:= Solve[Subscript[C, 1] == 0, En]  
Out[14]= $\left\{ En \to -\frac{c \, \omega^2 \, \hbar \, n_\theta^2}{a Z \alpha (1 - \epsilon^2)} \right\}$

Alternatively, from $(\ref{BidSom17})$:

In[15]:= Solve[(En Z \[Alpha])/(\!\(
\*SubsuperscriptBox[\(n\), \(\[Theta]\), \(2\)]*c*\[HBar]*
\*SuperscriptBox[\(\[Omega]\), \(2\)]\)) == 1/(a (1 - \[Epsilon]^2)), En]
Out[15]= $\left\{ En \to -\frac{c \, \omega^2 \, \hbar \, n_\theta^2}{a Z \alpha (1 - \epsilon^2)} \right\}$

We verify that the linear term is vanished:

In[16]:= Subscript[C, 1] /. En -> -((c \[Omega]^2 \[HBar] 
\!\(\*SubsuperscriptBox[\(n\), \(\[Theta]\), \(2\)]\))/(
   a Z \[Alpha] (-1 + \[Epsilon]^2)))  
Out[16]= $0$

The constant term can be simplified in terms of $(\ref{BidSom16})$:

In[17]:= Subscript[C, 2] /. 
 1/(-1 + \[Epsilon]^2) -> -((Subscript[n, r] + Sqrt[-Z^2 \[Alpha]^2 + 
\!\(\*SubsuperscriptBox[\(n\), \(\[Theta]\), \(2\)]\)])^2/(\[Omega]^2 
\!\(\*SubsuperscriptBox[\(n\), \(\[Theta]\), \(2\)]\)))
Out[17]= $\displaystyle -1 + \frac{En^2}{c^4 m^2} + \frac{\hbar^2 (n_r + \sqrt{-Z^2 \alpha^2 + n_\theta^2})^2}{a^2 c^2 m^2}$

The same for the linear term, in order to eliminate epsilon from the expression for energy:

In[18]:= Subscript[C, 1] /. 
 1/(-1 + \[Epsilon]^2) -> -((Subscript[n, r] + Sqrt[-Z^2 \[Alpha]^2 + 
\!\(\*SubsuperscriptBox[\(n\), \(\[Theta]\), \(2\)]\)])^2/(\[Omega]^2 
\!\(\*SubsuperscriptBox[\(n\), \(\[Theta]\), \(2\)]\)))
Out[18]= $\displaystyle \frac{2 En Z \alpha \hbar}{c^3 m^2} - \frac{2 \hbar^2 (n_r + \sqrt{-Z^2 \alpha^2 + n_\theta^2})^2}{a c^2 m^2}$

In[19]:= Solve[% == 0, En]
Out[19]= $\displaystyle \left\{ En \to \frac{c \hbar (n_r + \sqrt{-Z^2 \alpha^2 + n_\theta^2})^2}{a Z \alpha} \right\}$

Exclude energy from the constant term:

In[20]:= -1 + En^2/(c^4 m^2) + (\[HBar]^2 (Subscript[n, r] + Sqrt[-Z^2 \[Alpha]^2 + 
\!\(\*SubsuperscriptBox[\(n\), \(\[Theta]\), \(2\)]\)])^2)/(a^2 c^2 m^2) /. 
 En -> (c \[HBar] (Subscript[n, r] + Sqrt[-Z^2 \[Alpha]^2 + 
\!\(\*SubsuperscriptBox[\(n\), \(\[Theta]\), \(2\)]\)])^2)/(a Z \[Alpha])
Out[20]= $\displaystyle -1 + \frac{\hbar^2 \left(n_r + \sqrt{-Z^2 \alpha^2 + n_\theta^2}\right)^2}{a^2 c^2 m^2} + \frac{\hbar^2 \left(n_r + \sqrt{-Z^2 \alpha^2 + n_\theta^2}\right)^4}{a^2 c^2 m^2 Z^2 \alpha^2}$

Now solve a modified equation for $A = a^2$:

In[21]:= Solve[(\[HBar]^2 (Subscript[n, r] + Sqrt[-Z^2 \[Alpha]^2 + 
\!\(\*SubsuperscriptBox[\(n\), \(\[Theta]\), \(2\)]\)])^2)/(
   c^2 m^2) + (\[HBar]^2 (Subscript[n, r] + Sqrt[-Z^2 \[Alpha]^2 + 
\!\(\*SubsuperscriptBox[\(n\), \(\[Theta]\), \(2\)]\)])^4)/(
   c^2 m^2 Z^2 \[Alpha]^2) == A, A]
Out[21]= $\left\{ 
A \to \frac{\hbar^2 \left(n_r + \sqrt{-Z^2 \alpha^2 + n_\theta^2}\right)^2}{c^2 m^2} + 
\frac{\hbar^2 \left(n_r + \sqrt{-Z^2 \alpha^2 + n_\theta^2}\right)^4}{c^2 m^2 Z^2 \alpha^2} 
\right\}$

Transform the result as follows:

In[22]:= Factor[(\[HBar]^2 (Subscript[n, r] + Sqrt[-Z^2 \[Alpha]^2 + 
\!\(\*SubsuperscriptBox[\(n\), \(\[Theta]\), \(2\)]\)])^2)/(
  c^2 m^2) + (\[HBar]^2 (Subscript[n, r] + Sqrt[-Z^2 \[Alpha]^2 + 
\!\(\*SubsuperscriptBox[\(n\), \(\[Theta]\), \(2\)]\)])^4)/(
  c^2 m^2 Z^2 \[Alpha]^2)]
Out[22]= $\displaystyle \frac{\hbar^2 \left(n_r + \sqrt{-Z^2 \alpha^2 + n_\theta^2} \right)^2 \left(n_r^2 + n_\theta^2 + 2 n_r \sqrt{-Z^2 \alpha^2 + n_\theta^2} \right)}{c^2 m^2 Z^2 \alpha^2}$

An equivalent expression:

In[23]:= (\[HBar]^2 (Subscript[n, r] + Sqrt[-Z^2 \[Alpha]^2 + 
\!\(\*SubsuperscriptBox[\(n\), \(\[Theta]\), \(2\)]\)])^2*((Subscript[n, r] + 
     Sqrt[-Z^2 \[Alpha]^2 + 
\!\(\*SubsuperscriptBox[\(n\), \(\[Theta]\), \(2\)]\)])^2 + 
   Z^2 \[Alpha]^2))/(c^2 m^2 Z^2 \[Alpha]^2)
Out[23]= $\displaystyle \frac{\hbar^2 \left(n_r + \sqrt{-Z^2 \alpha^2 + n_\theta^2} \right)^2 \left( \left(n_r + \sqrt{-Z^2 \alpha^2 + n_\theta^2} \right)^2 + Z^2 \alpha^2 \right)}{c^2 m^2 Z^2 \alpha^2}$

Verification:

In[24]:= Simplify[% - %%]
Out[24]= 0

Finally, substitute the parameter $a$ into the previously found expression for energy:

In[25]:= (c \[HBar] (Subscript[n, r] + Sqrt[-Z^2 \[Alpha]^2 + 
\!\(\*SubsuperscriptBox[\(n\), \(\[Theta]\), \(2\)]\)])^2)/(a Z \[Alpha]) /. 
 a -> (\[HBar] (Subscript[n, r] + Sqrt[-Z^2 \[Alpha]^2 + 
\!\(\*SubsuperscriptBox[\(n\), \(\[Theta]\), \(2\)]\)]) Sqrt[
   Z^2 \[Alpha]^2 + (Subscript[n, r] + Sqrt[-Z^2 \[Alpha]^2 + 
\!\(\*SubsuperscriptBox[\(n\), \(\[Theta]\), \(2\)]\)])^2])/(c m Z \[Alpha])
Out[25]= $\displaystyle \frac{c^2 m \left(n_r + \sqrt{-Z^2 \alpha^2 + n_\theta^2} \right)}{\sqrt{Z^2 \alpha^2 + \left(n_r + \sqrt{-Z^2 \alpha^2 + n_\theta^2} \right)^2}}$

Compare the square of this result with the square of equation $(\ref{BidSomEnd})$:

In[26]:= Simplify[(Z^2 \[Alpha]^2 + (Subscript[n, r] + Sqrt[-Z^2 \[Alpha]^2 + 
\!\(\*SubsuperscriptBox[\(n\), \(\[Theta]\), \(2\)]\)])^2)/(Subscript[n, r] + 
     Sqrt[-Z^2 \[Alpha]^2 + 
\!\(\*SubsuperscriptBox[\(n\), \(\[Theta]\), \(2\)]\)])^2 - (1 + (
    Z^2 \[Alpha]^2)/(Subscript[n, r] + Sqrt[-Z^2 \[Alpha]^2 + 
\!\(\*SubsuperscriptBox[\(n\), \(\[Theta]\), \(2\)]\)])^2)]
Out[26]= 0

Thus we derived equations $(\ref{BidSomEnd})-(\ref{BidSom20})$ with the help of Mathematica computer algebra system.
\end{mat}

End of Mathematica session. (More details can be found in Ref.~\cite{BarleySusMath}.)

\medskip
%\vfill\eject\newpage
\newpage

\end{document}